\documentclass[a4paper,12pt]{article}
\usepackage[latin1]{inputenc}
\usepackage[T1]{fontenc}
\usepackage[english]{babel}
\usepackage{amssymb,amsbsy,amsmath,float,epsfig,here,amsthm}
\usepackage{babel,indentfirst}
\usepackage{graphicx}
\usepackage{geometry,color}
\usepackage{fancyhdr}
\usepackage{tabularx}
\usepackage{subfigure}
\usepackage{setspace}

\newtheorem{theorem}{Theorem}[section]
\newtheorem{proposition}{Proposition}[section]
\newtheorem{lemma}{Lemme}[section]

\DeclareMathOperator{\sech}{sech} 
\DeclareMathOperator{\diag}{diag}

\title{Semi-classical signal analysis}

\author{Taous-Meriem Laleg Kirati\thanks{INRIA, Bordeaux sud Ouest, Universit\'e de Pau et des Pays de l'Adour, UFR Sciences, B\^{a}t B1, BP 1155, 64013, Pau cedex, ({\tt Taous-Meriem.Laleg@inria.fr}).}
        \and Emmanuelle Cr\'epeau\thanks{Versailles Saint Quentin en Yvelines University, B\^{a}t Fermat, 78035, Versailles cedex,  ({\tt emmanuelle.crepeau@math.uvsq.fr}).}
        \and Michel Sorine\thanks{INRIA, centre Paris-Rocquencourt, Domaine de Voluceau, BP 105, 78153 Le Chesnay cedex, ({\tt Michel.Sorine@inria.fr}).}}

\date{}
\begin{document}

\maketitle

\paragraph{Abstract:}
This study introduces a new signal analysis method called SCSA, based on a semi-classical approach.
The main idea in the SCSA is to interpret a pulse-shaped signal as a potential of a Schr\"{o}dinger operator and then to use the discrete spectrum of
this operator for the analysis of the signal. We present some numerical examples and the first results obtained with this method on the analysis of arterial blood pressure waveforms.


\paragraph{AMS:}
00A69, 94A12, 92C55

%\pagestyle{myheadings}
%\thispagestyle{plain}
%\markboth{T.M. LALEG-KIRATI, E. CR\'{E}PEAU and M. SORINE}{SEMI-CLASSICAL SIGNAL ANALYSIS}

\section{Introduction}
This paper considers a new signal analysis method where the main idea is to interpret a signal $y$ as a multiplication operator, $\phi \rightarrow y.\phi$,
on some function space. The spectrum of a (formally) regularized version of this operator, denoted $H_h(y)$ and defined by
\begin{equation}\label{schrchap2}
H_h(y)\psi  = -h^2\frac{d^2\psi}{dx^2} - y\psi,  \quad  \psi\in  H^2(\mathbb{R}),\quad h>0,
 \end{equation}
for small $h$, is then used for the analysis instead of the Fourier transform of $y$. Here $H^2(\mathbb{R})$ denotes the Sobolev space of order 2.
In this method, the signal is  interpreted as a potential of a  Schr\"{o}dinger operator. This point of view seems useful when associated inverse spectral problem is well posed as it will be the case  for some pulse shaped signals.

We define $H_h$ on a space $\mathcal{B}$ such that
\begin{equation}\label{hypotheses}
\mathcal{B}=\{y \in L_1^1(\mathbb{R}), \quad y(x)\geq 0,\quad  \forall x\in
\mathbb{R},\quad \frac{\partial^m y}{\partial x^m}\in
L^1(\mathbb{R}), \quad m=1,2\},
\end{equation}
with,
$\displaystyle{L_1^1(\mathbb{R})=
\{V | \int_{-\infty}^{+\infty}{|V(x)|(1+|x|) dx}<\infty\}}$.  $L_1^1(\mathbb{R})$ is known as the Faddeev class \cite{Fad:64}.

For $\lambda \leq 0$, we denote $ N_h(\lambda;y)$ the number of eigenvalues of the
operator $H_h(y)$ below $\lambda$. Under hypothesis (\ref{hypotheses}), there is a non-zero finite number $N_h=N_h(0;y)$, as it is described in proposition \ref{nombre de composantes}. We denote $-\kappa_{nh}^2$ the negative
eigenvalues of $H_h(y)$ with $\kappa_{nh}
> 0$ and $\kappa_{1h}> \kappa_{2h}> \cdots > \kappa_{nh}$,
$n=1,\cdots,N_h$. Let $\psi_{nh}$, $n=1,\cdots, N_h$ be
the associated $L^2$-normalized eigenfunctions.

In this study, we focus our interest in representing the signal $y$ with the discrete spectrum of $H_h(y)$ using the following formula (\ref{formule introduction})
\begin{equation}\label{formule introduction}
y_h(x)= 4 h \sum_{n=1}^{N_h}{\kappa_{nh
}\psi_{nh}^2(x)}, \quad \quad x\in \mathbb{R}.
\end{equation}
As we will see, the parameter $h$ plays an important role in our approach. Indeed,  as $h$ decreases as the approximation of $y$ by $y_h$ improves.
Our method  is based on semi-classical concepts i.e $h\rightarrow 0$ and we will call it Semi-Classical Signal Analysis (SCSA).

%Supposing that  $-y$ has a minimum $-y_{max}$ (see figure \ref{puist_introduction}) and $\lambda$ is such that $ -y_{max} < \lambda \leq 0$, then there are two values $x_-$ and  $x_+$ for which
%\begin{equation}
%    y(x_-) = y(x_+)= \lambda, \quad \quad x_- < x_+.
%\end{equation}
%According to the Weyl law on the eigenvalues density \cite{HeRo:90a},  $ N_h(\lambda;y)$ is given by
%\begin{equation}\label{weyl2}
%    N_h (\lambda,y) \simeq \frac{1}{\pi h} \Phi(\lambda), \quad
%    \quad\mbox{with} \quad\quad
%     \Phi(\lambda)= \int_{x_-}^{x_+}{\sqrt{y(x) + \lambda}dx} .
%\end{equation}
%
%\begin{figure}[htbp]
%  \begin{center}
%  \includegraphics[width=8cm]{Figures/puits_introduction.eps}\\
%  \caption{A mono-well negative potentiel}\label{puist_introduction}\end{center}
%\end{figure}
%

In the next section, we will present some properties of the SCSA. In section 3, we  will consider a particular case of an exact representation for a fixed $h$ and show its relation to
a signal representation using the so called reflectionless potentials of the Schr\"{o}dinger operator.
Section 4 will deal with some numerical examples and section 5 will present some results obtained on the analysis of Arterial Blood Pressure (ABP) signals using the SCSA.
A discussion will summarize the main results and compare the SCSA  to related studies. In appendices A, B and C some known results on direct and inverse scattering transforms are presented.

%------------------------------------------------------------------------------------------------
\section{SCSA properties}
%-----------------------------------------------------------------------------------------------------------

To begin, we focuss our attention on the behavior of the number $N_h$ of negative
eigenvalues of $H_h(y)$ according to $h$ as described by the following proposition.

\begin{proposition}\label{nombre de composantes}
Let $y$ be a real valued function satisfying hypothesis
(\ref{hypotheses}). Then,
\begin{itemize}
\item[i) ] The number $N_h$ of negative eigenvalues of $H_h(y)$ is a decreasing function of $h$.\\

\item[ii)] Moreover if  $y\in L^{\frac{1}{2}}(\mathbb{R})$, then,
\begin{equation}\label{comportement asymptotique de Nchi1}
    \lim_{h\rightarrow 0}{hN_h} = \frac{1}{\pi} \int_{-\infty}^{+\infty}{\sqrt{y(x)}dx},
\end{equation}
\end{itemize}
\end{proposition}

{\em Proof}.
\begin{itemize}
\item [i)] The proof of this item is based on the following lemma.

\begin{lemma}\cite{BlSt:96}\cite{ReSi:78} \label{lemmeN}
For $\lambda \leq 0$, let $N_1(\lambda;V)$ be the number of
negative eigenvalues of $H_1(V)$ less than $\lambda$. Let $V$ and
$W$ be two potentials of the Schr\"{o}dinger operator such that $W
\leq V$, then
\begin{equation}
N_1(\lambda;W)\leq N_1(\lambda;V), \quad \forall \lambda\leq 0.
\end{equation}
\end{lemma}
Let $y\in \mathcal{B}$. We put
$\displaystyle{
V =\frac{1}{h_1^2} y, \quad \quad  W = \frac{1}{h_2^2} y, \quad\quad \mbox{with } \quad
0< h_1 \leq h_2}$.

For $\lambda=0$, we obtain
$\displaystyle{
N_1(0;\frac{1}{h_2^2} y) \leq N_1(0;\frac{1}{h_1^2} y),}$
and we have $N_1(0;\frac{1}{h_j^2} y) = N_{h_{j}}$, $j=1,2$, which proves the result.\\

\item [ii)]
Let $y\in \mathcal{B} \cap L^{\frac{1}{2}}(\mathbb{R}) $ and $\lambda\in
]-y_{max},0]$.  We denote
\begin{equation}\label{moyennes de riesz}
S_\gamma(h,\lambda) = \sum_{\kappa_{nh}^2\leq
\lambda}{\left(\lambda +
\kappa_{nh}^2\right)^\gamma},\quad\quad \gamma
\geq 0
\end{equation}
the Riesz means of the values $-\kappa_{nh}^2$
less than $\lambda$. Remark that  $S_0(h,\lambda) = N_h(\lambda;y) $.

Property ii) results from the following lemma \ref{formule
asymptotique de weyl}
\begin{lemma}\label{formule asymptotique de weyl}\cite{LaWe:00}
For $y\in L^{\gamma+\frac{1}{2}}(\mathbb{R})$, $y(x)\geq 0$,
$\forall x\in \mathbb{R}$ and $\gamma \geq 0$, we have
\begin{equation}\label{formule asymptotique de weyl equation}
\lim_{h\rightarrow 0}{h S_\gamma(h,0)} = L_{\gamma}^{cl}\int_{-\infty}^{+\infty}{y(x)^{\gamma+\frac{1}{2}}dx},
\end{equation}
where  $L_{\gamma}^{cl}$ is the classical constant given by
\begin{equation}\label{theoreme_moyenne de Riesz equation3}
   L_{\gamma}^{cl}=\frac{\Gamma(\gamma+1)}{2\sqrt{\pi}
     \Gamma(\gamma+\frac{3}{2})},\end{equation}
for all $\gamma\geq
0$
\end{lemma}

By taking $\gamma=0$ in (\ref{formule
asymptotique de weyl equation}) we get the result.

\end{itemize}
\endproof

Let us now study some properties of the negative eigenvalues
$-\kappa_{nh}^2$, $n=1,\cdots,N_h$ of $H_h(y)$.
\begin{proposition}\label{comportement asymptotique VAP}
Let $y\in C^\infty(\mathbb{R})$, with $y(x) >0$, $\forall x\in \mathbb{R}$ and such that $\exists \gamma_0 \in \mathbb{R}$, $\min_{\mathbb{R}}{(-y+\gamma_0)} >
0$ and  $\forall \alpha \in \mathbb{N}$, $\exists C_\alpha > 0$ such that $|\frac{\partial^\alpha y}{\partial
x^\alpha}| \leq C_\alpha (-y+\gamma_0)$, then every regular value
of  $y$ is an accumulation point of the set
($\kappa_{nh}^2$, $n=1,\cdots,N_h$)
($v$ is a regular value if $0 < v < y_{max}$ and if $y(x) = v$
then $|\frac{dy(x)}{dx}| > 0$).\\
\end{proposition}

{\em Proof}.

We want to show that every regular value of $y$ is an
accumulation point for the set ($\kappa_{nh}^2$, $n=1,\cdots,N_h$). For this purpose, we use the following result shown by B. Helffer and D. Robert \cite{HeRo:90a}.
\begin{theorem}\label{theoreme helffer robert}\cite{HeRo:90a}
Let $y\in C^\infty(\mathbb{R})$, with $y(x)>0$, $\forall x \in
\mathbb{R}$ such that for $\gamma_0 \in \mathbb{R}$,
$\min_{\mathbb{R}}{(-y+\gamma_0)} > 0$ and for all $\alpha \in
\mathbb{N}$, there is a constant $C_\alpha
> 0$ such that $|\frac{\partial^\alpha y}{\partial x^\alpha}| \leq
C_\alpha (-y+\gamma_0)$.  Let $\lambda < \lim \inf_{|x|\rightarrow\infty} ({- y(x)})$, then for $0\leq \gamma \leq 1$ the Riesz means (\ref{moyennes de riesz}) are given by
\begin{equation}\label{theoreme_moyenne de Riesz equation1}
    S_\gamma(h, \lambda)= \frac{1}{h}\left(L_\gamma^{cl}\int_{-\infty}^{+\infty}{\left|\lambda +
y(x)\right|_+ ^{\gamma +\frac{1}{2}} dx} +
    O(h^{1+\gamma})\right),
\end{equation}
where $|V|_+$ is the positive part of $V$ and $L_\gamma^{cl}$,
known as the classical constant, is given by (\ref{theoreme_moyenne de Riesz equation3}).\\
     \end{theorem}

We put $\gamma=0$ in  (\ref{moyennes de riesz}). We notice that $S_0(h,\lambda) = N_h(\lambda,y)$. Substituting $\gamma$ by $0$  in (\ref{theoreme_moyenne
de Riesz equation1}), we get
\begin{equation}\label{preuve1}
    S_0(h, \lambda)= \frac{1}{h}L_0^{cl}\int_{-\infty}^{+\infty}{\sqrt{\left|\lambda +
y(x)\right|_+} dx} + O(1).
\end{equation}

We suppose that there is a regular value $y_0$ of $y$ that is not
an accumulation point of the set ($\kappa_{nh}^2$, $n=1,\cdots,N_h$). So there is a neighborhood
$V(y_0)$ of $y_0$ and a value $h_0$, small enough such that
$\forall h < h_0$, $V(y_0)$ does not
contain any element element $\kappa_{nh}^2$.\\

Moreover, we can choose $V(y_0)$ small enough such that
\begin{equation}
    \inf\{|\frac{d y(x)}{dx}|, \quad \mbox{for} \quad x \quad \mbox{such that} \quad  y(x) \in V(y_0)
    \}= c > 0.
\end{equation}
Then, we can take $V(y_0)= [y_1,y_2[$, with $y_1$ and $y_2$ some
regular values of $y$ and $ 0 < y_1 < y_2$.\\

For all $h < h_0$, the difference $S_0(h, -y_1) -
S_0(h, -y_2)$ represents the number of elements of
($-\kappa_{nh}^2$, $n=1,\cdots,N_h$)
in the interval $]-y_2,-y_1]$. However this set is empty because there is no element
in the neighborhood of $y_0$.\\

Denoting $X(\lambda)= \{x|y(x)+\lambda \geq 0\}$, we have from
(\ref{preuve1})
\begin{eqnarray}\label{preuve3}
\!\!\!    S_0(h,-y_1)\!-\!S_0(h,-y_2) &=&
\frac{1}{h}L_0^{cl}\!\left(\int_{X(-y_1)}
{\!\!\!\!\!\!\!\!\sqrt{y(x) - y_1}
dx}\!-\!\int_{X(-y_2)}{\!\!\!\!\!\!\!\!\sqrt{y(x) - y_2}
dx}\right)\!\! \nonumber\\&+& \!\! O(1),\!
\end{eqnarray}
so, as the left quantity is null, we obtain
\begin{equation}
\int_{X(-y_1)}{\sqrt{y(x) - y_1}dx}= \int_{X(-y_2)}{\sqrt{y(x) -
y_2} dx},
\end{equation}
and we have  $X(-y_1)= X(-y_2) \cup y^{-1}([y_1,y_2[)$, so
\begin{equation}
\int_{X(-y_2)}{\!\!\!\!\!\!\!\!\!\sqrt{y(x) - y_1}dx}+
\int_{y^{-1}([y_1,y_2[)}{\!\!\!\!\!\!\!\!\!\sqrt{y(x) - y_1} dx}=
\int_{X(-y_2)}{\!\!\!\!\!\!\!\!\!\sqrt{y(x) - y_2} dx},
\end{equation}
\begin{equation}
\int_{X(-y_2)}{(\sqrt{y(x) - y_1} - \sqrt{y(x) - y_2}) dx}+
\int_{y^{-1}([y_1,y_2[)}{\sqrt{y(x) - y_1} dx}=0.
\end{equation}
Hence as these two integrals are positives, we get
\begin{equation}
\int_{y^{-1}([y_1,y_2[)}{\sqrt{y(x) - y_1} dx}=0,
\end{equation}
therefore $|y(x) -y_1| = 0$ almost everywhere in $y^{-1}([y_1,
y_2[)$, then $y_1=y_2$, which is a contradiction.

\endproof

\begin{figure}[htbp]
  \begin{center}
  \includegraphics[width=8cm]{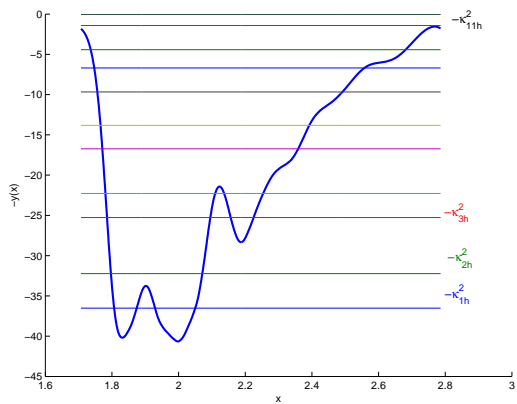}\\
  \caption{The negative spectrum of the Schr\"{o}dinger operator $-h^2\dfrac{d ^2 }{d x^2} - y(x)$
   provides a quantization  of the signal $y$}\label{PA_VAP}\end{center}
\end{figure}

Now, proposition \ref{comportement asymptotique VAP} introduces an interesting property of the SCSA. Indeed, remembering that
\begin{equation}
    -h^2 \frac{d^2 \psi_{nh}(x)}{d x^2}- y(x) \psi_{nh}(x) = -\kappa_{nh}^2 \psi_{nh}(x).
\end{equation}
and by multiplying the previous equation by $\psi_{nh}(x)$ and integrating it by part we get
\begin{equation}\label{eq3 chapitre3}
\kappa_{nh}^2= \int_{-\infty}^{+\infty}{y
\psi_{nh}^2(x)dx} - h^2\int_{-\infty}^{+\infty}{\left(\frac{d \psi_{nh}(x)}{d x}\right)^2 dx}.
\end{equation}
Then, we notice that $- y_{max} \leq -\kappa_{nh}^2 < 0$ as it is illustrated in figure
\ref{PA_VAP}. Hence, for a fixed value of $h$,  $\kappa_{nh}^2$  can be interpreted as particular values of $y$ which define a new quantization approach that can be interpreted by semi-classical concepts. The SCSA appears then as a new way to quantify a signal.

%Indeed, in a semi-classical regime, i.e when $h\rightarrow 0$, the possible values for $-\kappa_{nh}^2 \leq \lambda $ are such that $\Phi(\lambda)$ is an integer (à un demi pres) multiple of $\pi h$. This property is given by the Bohr-Sommerfeld condition:
%\begin{equation}\label{bohr-sommerfeld}
%    \Phi(\lambda)= (n + \frac{1}{2}) \pi h.
%\end{equation}
%Therefore, a signal, taken as a potentiel of $H_h(y)$, can be represented by the values $\kappa_{nh}^2$, $n=1,\cdots,N_h$. We call this new signal analysis method Semi-Classical Signal Analysis (SCSA).

%\begin{figure}
%  % Requires \usepackage{graphicx}
% \begin{center}
% \includegraphics[width=12cm]{Figures/PA_VAP_Niveau}
%  \caption{The values $-\frac{\kappa_{n\chi}}{\chi}$ are all greater or equal to $-y_{max}$}\label{PA_VAP_Niveau}\end{center}
%\end{figure}

To finish this section, we examine the convergence of the first two momentums of
$\kappa_{nh}$  when $h\rightarrow 0$ through proposition \ref{convergence invariants chapitre3}. These quantities could be very interesting in signal analysis as
it is mentioned in section 5 and also in a recent study \cite{LaMePaCoVa:10}.

\begin{proposition}\label{convergence invariants chapitre3}
Under hypothesis (\ref{hypotheses}), we have
\begin{equation}\label{res1}
    \lim_{h\rightarrow 0}{h\sum_{n=1}^{N_h}{\kappa_{nh}}}=
    \frac{1}{4}\int_{-\infty}^{+\infty}{y(x)dx},
\end{equation}
and remembering that \begin{equation}
y_h(x)= 4 h \sum_{n=1}^{N_h}{\kappa_{nh
}\psi_{nh}^2(x)}, \quad \quad x\in \mathbb{R},
\end{equation}
we have
\begin{equation}\label{L1}
    \lim_{h\rightarrow 0}\int_{-\infty}^{+\infty}{y_h(x) dx}=\int_{-\infty}^{+\infty}{y(x)
dx}.
\end{equation}

Moreover if $y\in L^2(\mathbb{R})$, then,
\begin{equation}\label{res2}
    \lim_{h\rightarrow 0}{h\sum_{n=1}^{N_h}{\kappa_{nh}^3}}=
\frac{3}{16}\int_{-\infty}^{+\infty}{y^2(x)dx}.
\end{equation}\\
\end{proposition}

{\em Proof}.
The limits (\ref{res1}) and  (\ref{res2}) are deduced from lemma
\ref{formule asymptotique de weyl} for $\gamma=\dfrac{1}{2}$ and
$\gamma=\dfrac{3}{2}$ respectively,
\begin{equation}\label{res1_preuve}
    \lim_{h\rightarrow 0}{h\sum_{n=1}^{N_h}{\kappa_{nh}}}=
    \frac{1}{4}\int_{-\infty}^{+\infty}{y(x)dx},
    \end{equation}
\begin{equation}\label{res2_preuve}
    \lim_{h\rightarrow 0}{h\sum_{n=1}^{N_h}{\kappa_{nh}^3}}=
\frac{3}{16}\int_{-\infty}^{+\infty}{y^2(x)dx}.
\end{equation}

By integrating (\ref{formule introduction}) we have
%\begin{equation}
%    y_h(x)=h\sum_{n=1}^{N_h}{\kappa_{nh}
%\psi_{nh}^2(x)}\quad \quad \mbox{and} \quad \quad
%\int_{-\infty}^{+\infty}{\psi_{nh}^2(x) dx}=1,
%\end{equation}
\begin{equation}\label{integrale de y chi}
\int_{-\infty}^{+\infty}{y_h(x)
dx}= 4 h\sum_{n=1}^{N_h}{\kappa_{nh}}.
\end{equation}
Then, combining (\ref{integrale de y chi}) and (\ref{res1}), we
get (\ref{L1}).\\
\endproof

%--------------------------------------------------------------------------------------

\section{Exact representation and reflectionless potentials}
In this section, we are interested in an exact representation of a signal  for a fixed $h$  and its relation to reflectionless potentials
(reflectionless potentials are defined in the appendix B) of the Schr\"{o}dinger operator as it is described in the following proposition:
\begin{proposition}\label{approximation exacte}
The following properties are equivalent,
\begin{itemize}
\item [i)]  Equality in (\ref{res1}) holds for a finite $h$ ;
\item [ii)] $\exists h$ such that $y_h = y$ ;
\item [iii)]$\exists h$ such that $ \dfrac{y}{h^2}$ is a reflectionless potential of $H_1(V)$.
\end{itemize}

\end{proposition}

{\em Proof}

The following proof uses some concepts and results from scattering transform theory that are recalled in the appendix.
%In particular, we recall the definition of reflectionless potentials of the Schr\"{o}dinger operator.
First, we suppose that i) is fulfilled then
\begin{equation}\label{proposition3equation1}
 \exists h, \quad \int_{-\infty}^{+\infty}{y(x) dx}=4 h \sum_{n=1}^{N_h}{\kappa_{nh}}.
\end{equation}
Writing the first invariant \ref{firstinvariant} (see the appendix C) for the potential
$- \dfrac{y}{h^2}$
%and using $|T_\chi(k)|^2 = 1 - |R_{r(l)\chi}(k)|^2$,
we have
\begin{equation}\label{formule1}
 \int_{-\infty}^{+\infty}{y(x) dx}=4 h \sum_{n=1}^{N_h}{\kappa_{nh}}
 + \frac{h^2}{\pi } \int_{-\infty}^{+\infty}{\ln{(1-|R_{r(l)h}(k)|^2)}
dk},
\end{equation}
where $R_{r(l)h}(k)$ is the reflection coefficient (see the appendix A).

Using (\ref{integrale de y chi}), we get
\begin{equation}\label{formule3}
 \int_{-\infty}^{+\infty}{y(x) dx}=\int_{-\infty}^{+\infty}{y_h(x)
dx} + \frac{h^2}{\pi}
\int_{-\infty}^{+\infty}{\ln{(1-|R_{r(l)h}(k)|^2)} dk}.
\end{equation}
Then, from (\ref{formule0}) and (\ref{formule3}) we obtain
\begin{equation}\label{formule4}
\int_{-\infty}^{+\infty}{\ln{(1-|R_{r(l)h}(k)|^2)} dk}=0.
\end{equation}
The reflection coefficient of a Schr\"{o}dinger operator satisfies $|R_{r(l)h}(k)| \leq 1$, $\forall k\in
\mathbb{R}$ (see for exemple \cite{DeTr:79}). Then,  we get $\ln{(1-|R_{r(l)h}(k)|^2)} \leq 0$.
Equality (\ref{formule4}) is then fulfilled if and only if
$\ln{(1-|R_{r(l)h}(k)|^2)}=0$, $ k \in \mathbb{R} \:\:a.e$;
which is true only if $|R_{r(l)h}(k)|=0$, $ k \in
\mathbb{R} \:\:a.e$. This property defines a reflectionless potential. So, $i) \Rightarrow iii)$.

Now, using the Deift-Trubowitz formula (\ref{formule deift-trubowitz}) (see appendix B) that we rewrite  for the potential $-\dfrac{y}{h^2}$ and taking
 $R_{r(l)h}(k)=0$, we can deduce  that statement (iii) implies  statement (ii).

Then, if we suppose that $y_h=y$ for a
given value of  $h$ we have
\begin{equation}\label{formule0}
    \int_{-\infty}^{+\infty}{y_h(x) dx}=\int_{-\infty}^{+\infty}{y(x)
dx},
\end{equation}
hence $ii)\Rightarrow i)$
\endproof

%-----------------------------------------------------------------------------------------------------------

\section{Numerical results}
In this section, we are interested in the validation of the SCSA
through some numerical examples.
For this purpose, it will be more convenient to consider the problem associated to
$H_1(\dfrac{y}{h^2})$. Therefore, in order to  simplify the notations,
we put $\dfrac{1}{h^2}= \chi$, $N_h=N_\chi$ and $\dfrac{\kappa_{nh}^2}{h^2} = \kappa_{n\chi}^2$, $n=1,\cdots,N_\chi$.
We denote the $L^2$-normalized eigenfunctions $\psi_{n\chi}$, $n=1,\cdots,N_\chi$. Formula (\ref{formule introduction}) is then rewritten
\begin{equation}\label{formule chi}
y_\chi(x)= \frac{4}{\chi} \sum_{n=1}^{N_\chi}{\kappa_{n\chi
}\psi_{n\chi}^2(x)}, \quad \quad x\in \mathbb{R},
\end{equation}

We start by giving the numerical scheme used to estimate a signal with the SCSA. Then, the sech-squared function will be considered. This example illustrates the
influence of the parameter $\chi$ on the approximation. Gaussian, sinusoidal and chirp signals will be also considered.

\subsection{The numerical scheme}
The first step in the SCSA is to solve the spectral
problem of a one dimensional Schr\"{o}dinger operator. Its
discretization leads to an eigenvalue problem of a matrix. In this work, we propose to use
a Fourier pseudo-spectral method \cite{HuGoOr:84}, \cite{Tre:00}.
The latter is well-adapted for periodic problems but in practice it gives good results for some non-periodic problems for instance, rapid decreasing signals.

We consider a grid of $M$ equidistant points $x_j$, $j=1,\cdots,M$
such that
\begin{equation}
a=x_1<x_2<\cdots<x_{M-1}<x_M=b.
\end{equation}
Let $\Delta x=\dfrac{b-a}{M-1}$ be the distance between two consecutive
points. We denote $y_j$ and $\psi_j$ the values of $y$ and $\psi$ at the grid points
$x_j$, $j=1,\cdots, M$
\begin{equation}
y_j=y(x_j),\quad \psi_j=\psi(x_j),\quad j=1,\cdots, M.
\end{equation}

Therefore, the discretization of the Schr\"{o}dinger eigenvalue
problem leads to the following eigenvalue matrix problem
\begin{equation}\label{sch_discrete2}
    \left( -D_2- \chi \diag\left( Y \right) \right)\underline{\psi}=\-\lambda\underline{\psi},
\end{equation}
where $\diag(Y)$ is a diagonal matrix whose elements are $y_j$,
$j=1,\cdots,M$ and $\underline{\psi}= \left[\psi_1 \:\:
\psi_2,\:\:\cdots\:\: \psi_{M-1}\:\: \psi_M\right]^{T}$. $D_2$ is
the second order differentiation matrix given by \cite{Tre:00},
\begin{itemize}
\item  If  $M$ is even
\begin{equation}
D_2(k,j)=\frac{\Delta^2}{(\Delta x)^2}\left\{%
\begin{array}{ll}
  \dfrac{-\pi^2}{3\Delta^2}-\dfrac{1}{6} & \quad \mbox{for} \:\: k=j, \\
  -\left( -1 \right)^{k-j}\dfrac{1}{2}\dfrac{1}{\sin^2{\left( \frac{\left( k-j \right) \Delta}{2}\right)}} & \quad \mbox{for} \:\: k\neq j. \\
\end{array}%
\right. \end{equation}

\item If $M$ is odd
\begin{equation}
D_2(k,j)=\frac{\Delta^2}{(\Delta x)^2}\left\{%
\begin{array}{ll}
  \dfrac{-\pi^2}{3\Delta^2}-\dfrac{1}{12} & \quad \mbox{pour} \:\: k=j, \\
  -\left( -1 \right)^{k-j}\dfrac{1}{2}\dfrac{1}{\sin{\left( \frac{\left( k-j
  \right)\Delta}{2}\right)}}\cot{\left( \frac{\left( k-j \right)\Delta}{2} \right)}
  & \quad \mbox{pour} \:\: k\neq j, \\
\end{array}%
\right.
\end{equation}
\end{itemize}
with $\Delta=\dfrac{2\pi}{M}$.  the matrix $D_2$ is symmetric and
definite negative. To solve the eigenvalue problem of the matrix
$(-D_2-\chi \diag(Y))$ we use the Matlab routine eig.

\begin{figure}
  \begin{center}
  \includegraphics[width=8cm]{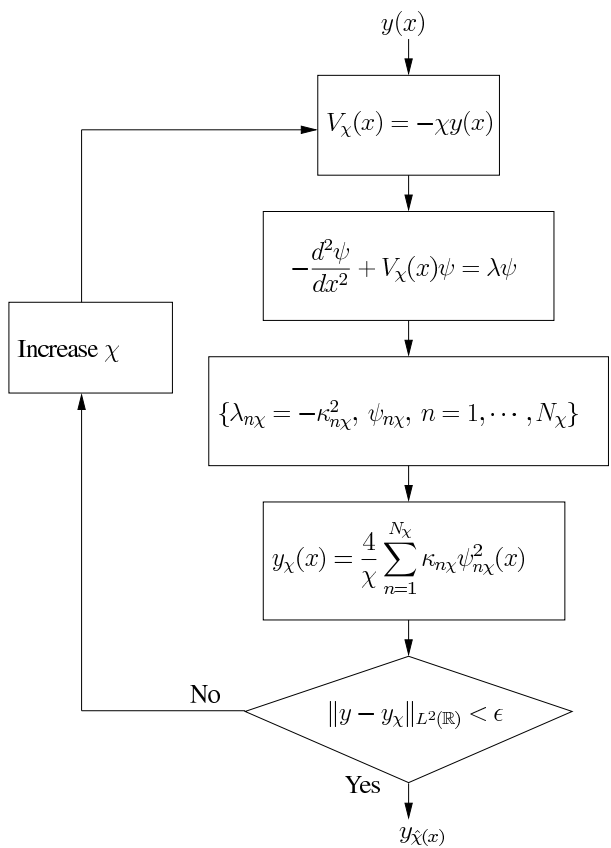}
  \caption{The SCSA algorithm}\label{principe}
  \end{center}
\end{figure}

The final step in the SCSA algorithm is to find an
optimal value of the parameter $\chi$. So we look for a value
$\hat{\chi}$ that gives a good approximation of $y$ with
a small number of negative eigenvalues. From the numerical tests, we noticed that the number $N_\chi$ is in general a step by step function of $\chi$. So, we optimize the following criteria in each interval
$\left[\chi_1,\chi_2\right] $ where $N_\chi$ is
constant,
\begin{equation}\label{critere}
    J(\chi)=\frac{1}{M}\sum_{i=1}^M{\left( y_i-y_{\chi i} \right)^2}, \quad y_{\chi
i}=\frac{4}{\chi}\sum_{n=1}^{N_\chi}{\kappa_{n\chi}\psi_{ni\chi}^2},
\quad i=1,\cdots,M.
\end{equation}
For $\chi$ large enough (equivalently $h$ small enough), we know an approximate  relation between
the number of negative eigenvalues and $\chi$ thanks to proposition \ref{nombre de composantes}.
%\begin{equation}
%N_\chi \sim \frac{1}{\pi}\int_{-\infty}^{+\infty}{\sqrt{|\chi
%y(x)|}dx}, \quad \chi \rightarrow +\infty.
%\end{equation}
Then, we can deduce approximate values of $\chi_1$ and $\chi_2$
according to a given number of negative eigenvalues.
Fig. \ref{principe} summarizes the SCSA algorithm.

\emph{\textbf{Remark:}} In practice, we often omit the
optimization step and just fix $N_\chi$ to a large enough
value.

\subsection{The sech-squared  function}
In order to illustrate the influence of the parameter $\chi$ on the SCSA,
we first study a sech-squared function given by
\begin{equation}\label{sech}
    y(x)=\sech^2(x-x_0), \quad x\in \mathbb{R}.
\end{equation}
The potential of the Schr\"{o}dinger operator $H_1(\chi y )$ is given in this case  by: $-\chi
\sech^2(x-x_0)$. This potential is called in quantum physics
\textit{P\"{o}schl-Teller} potential.

It is well-known that the \textit{P\"{o}schl-Teller}  potential belongs
to the class of reflectionless potentials if,
\begin{equation}\label{condition}
\chi=\chi_p=N(N+1),\quad \quad  N=1,2,3,\cdots,
\end{equation}
$N$ being  the number of negative eigenvalues of $H_1(\chi y)$
\cite{LaLi:58}.

So, for example, if $\chi=2$, the Schr\"{o}dinger operator spectrum is
negative and consists of a single negative eigenvalue given by
$\lambda=-1$. If $\chi=6$, there are two negative eigenvalues:
$\lambda_1=-4$, $\lambda_2=-1$ and so on.

Let us now apply the SCSA to reconstruct $y$. For
this purpose, we must truncate the signal and consider it on a
finite interval so that the numerical computations could be possible.

\begin{figure}[htbp]
\begin{center}
\subfigure{\epsfig{figure=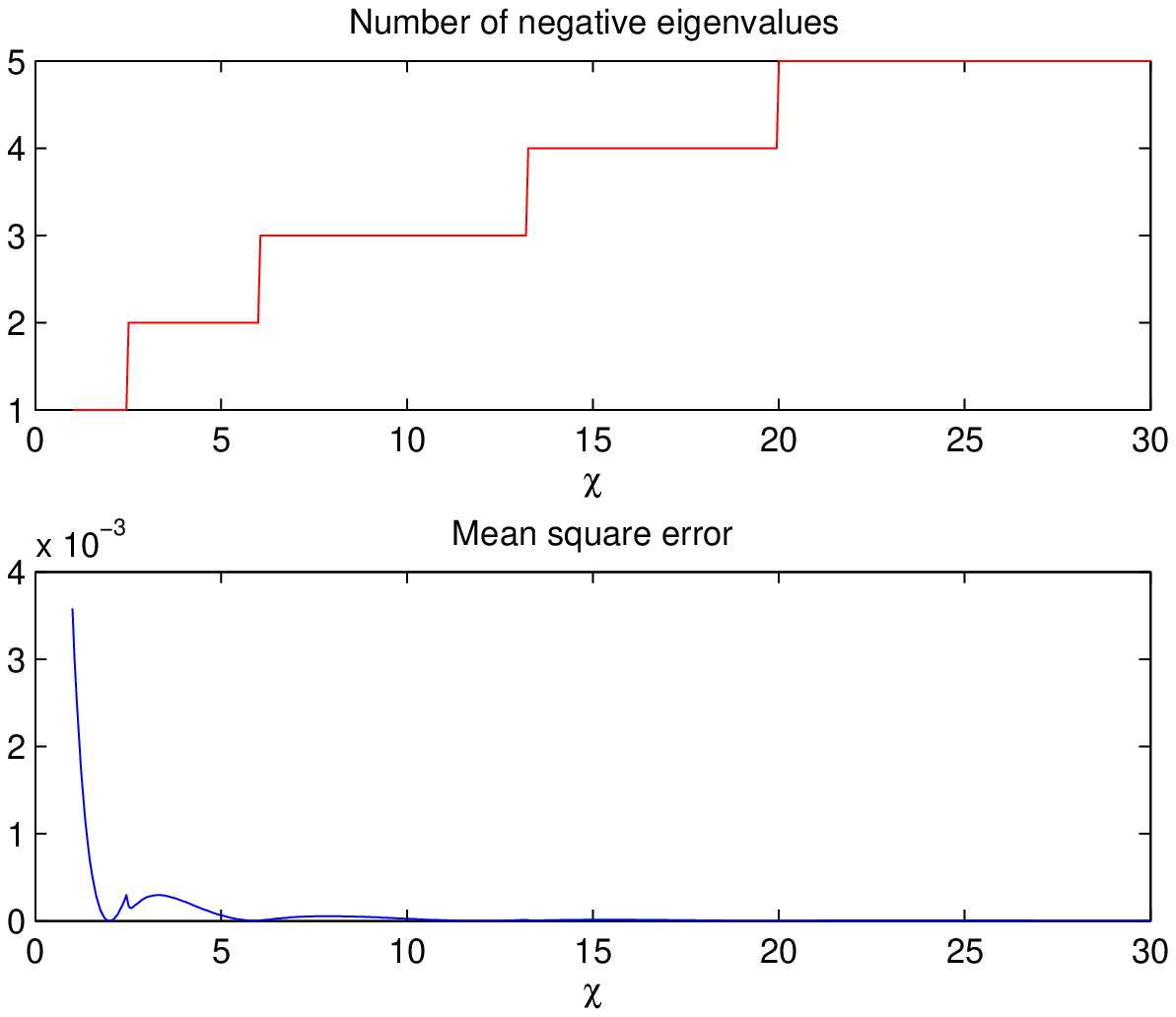,width=6cm}}
\subfigure{\epsfig{figure=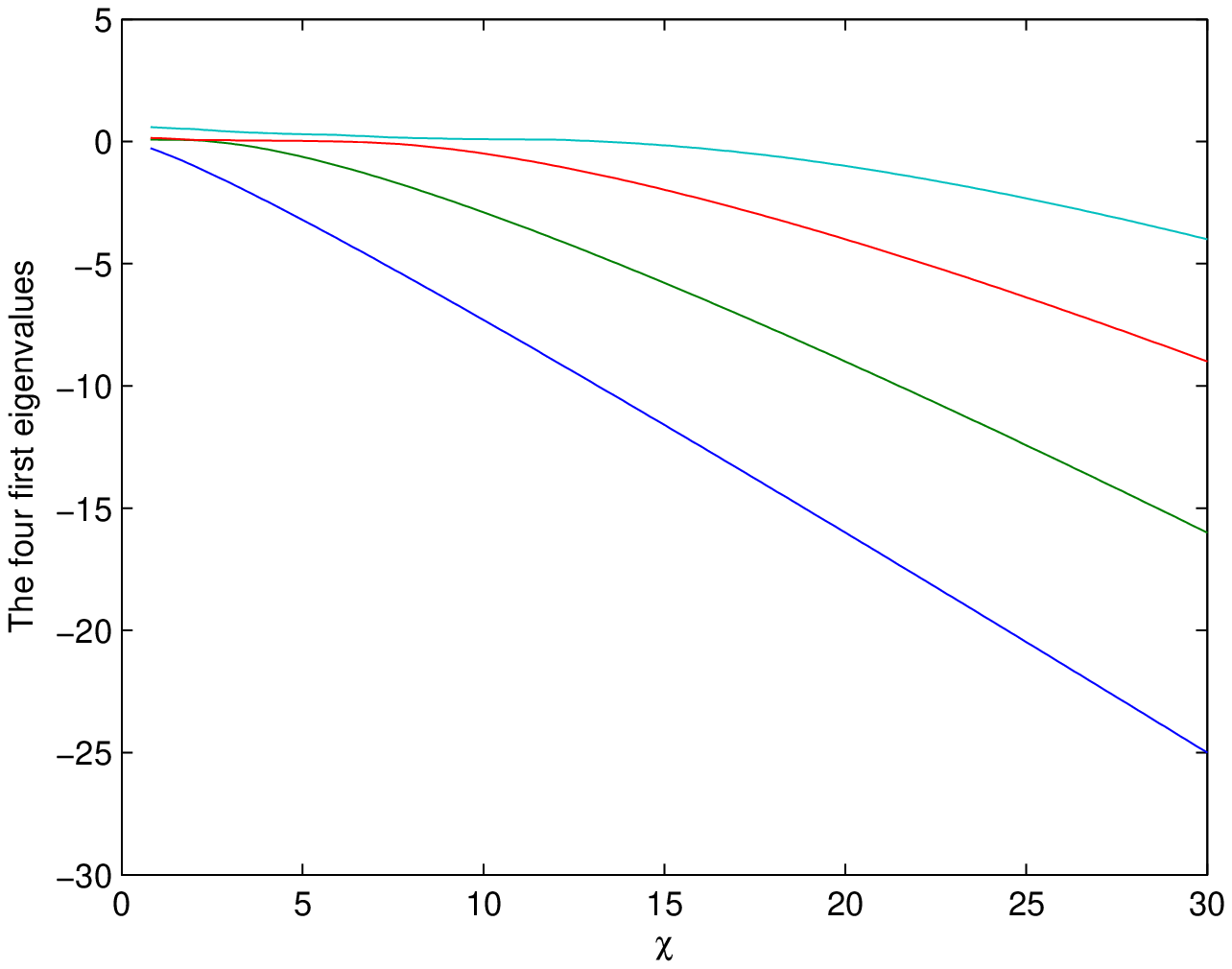,width=6cm}}\\
{\hspace{0.5cm} (a) \hspace{5cm} (b) }
\caption{(a) Mean square error and number of negative eigenvalues  according to $\chi$ for $y(x)=\sech^2(x-6)$ in
$\left[0,15\right]$.  (b) Four first eigenvalues according to  $\chi$ for $y(x)=\sech^2(x-6)$ in $\left[0,15\right]$} \label{erreur_sech2_SINC}
\end{center}
\end{figure}

Figure \ref{erreur_sech2_SINC}.a illustrates the variation of the
mean square error and $N_\chi$ according to $\chi$. We notice that,

\begin{itemize}
    \item $N_\chi$ is an increasing function of $\chi$ as described in proposition  (\ref{nombre de composantes}).
    Moreover, $N_\chi$ is a step by step function.\\

    \item There are some particular values of $\chi$ for which
the error is minimal. These values are in fact the particular
values $\chi_p=N_\chi(N_\chi+1)$, $N_\chi=1,2,\cdots$ for which
$-\chi y$ is a reflectionless potential.\\

    \item For all $\epsilon > 0$, there is a value  $\chi = \chi_\epsilon$ such that $\forall
    \chi > \chi_\epsilon, \quad J(\chi) < \epsilon$.\\
\end{itemize}

Figure \ref{erreur_sech2_SINC}.b illustrates the variation of the
 first four eigenvalues of the matrix $-D_2-\chi \diag(Y)$, settled in
an increasing way, according to $\chi$. We notice that these
eigenvalues, initially positive, are decreasing functions of
$\chi$ and at every passage from $N_\chi$ to $N_\chi+1$, a
positive eigenvalue becomes negative.

\begin{figure}[htbp]
\begin{center}
\subfigure{\epsfig{figure=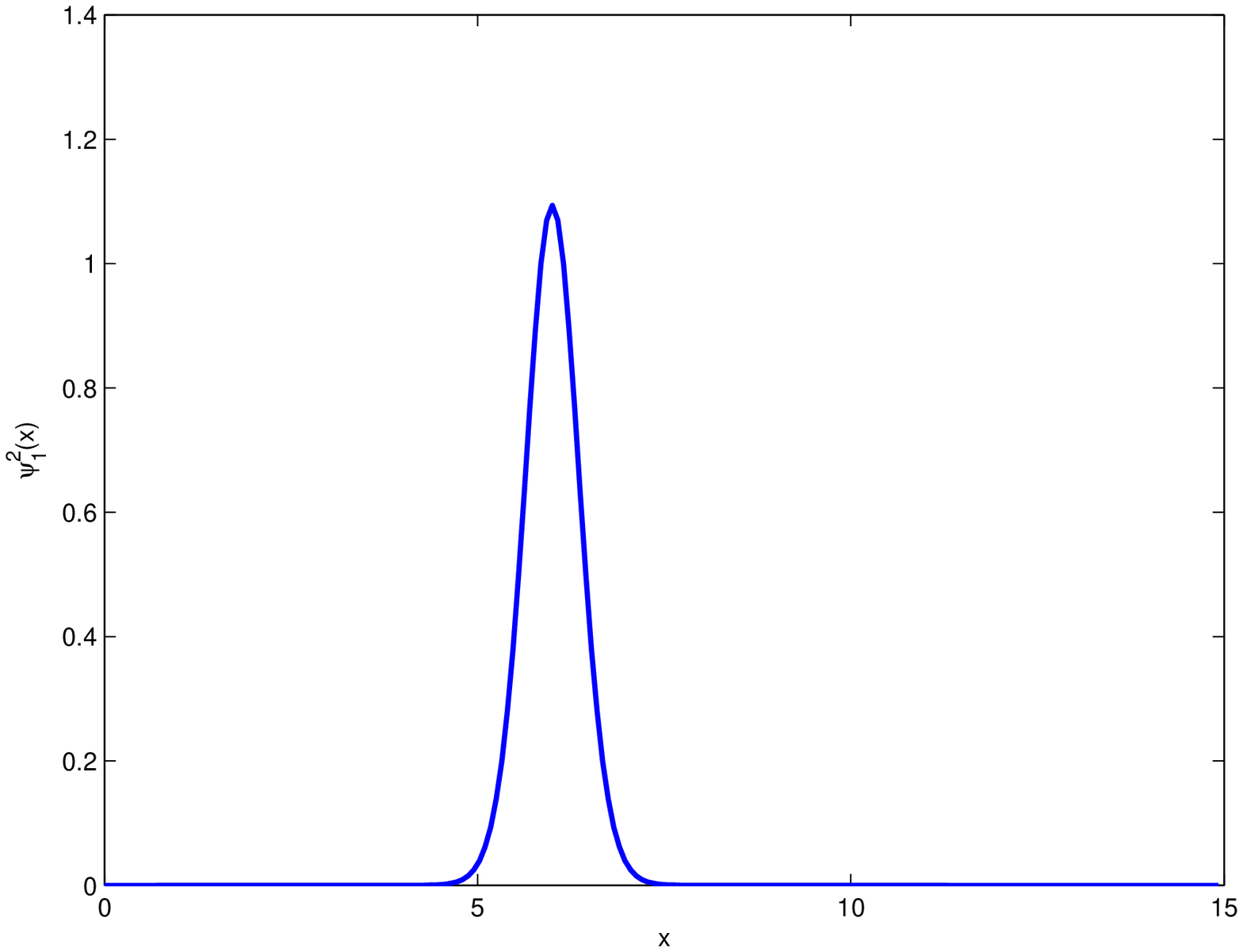,width=5.5cm}}\quad
\subfigure{\epsfig{figure=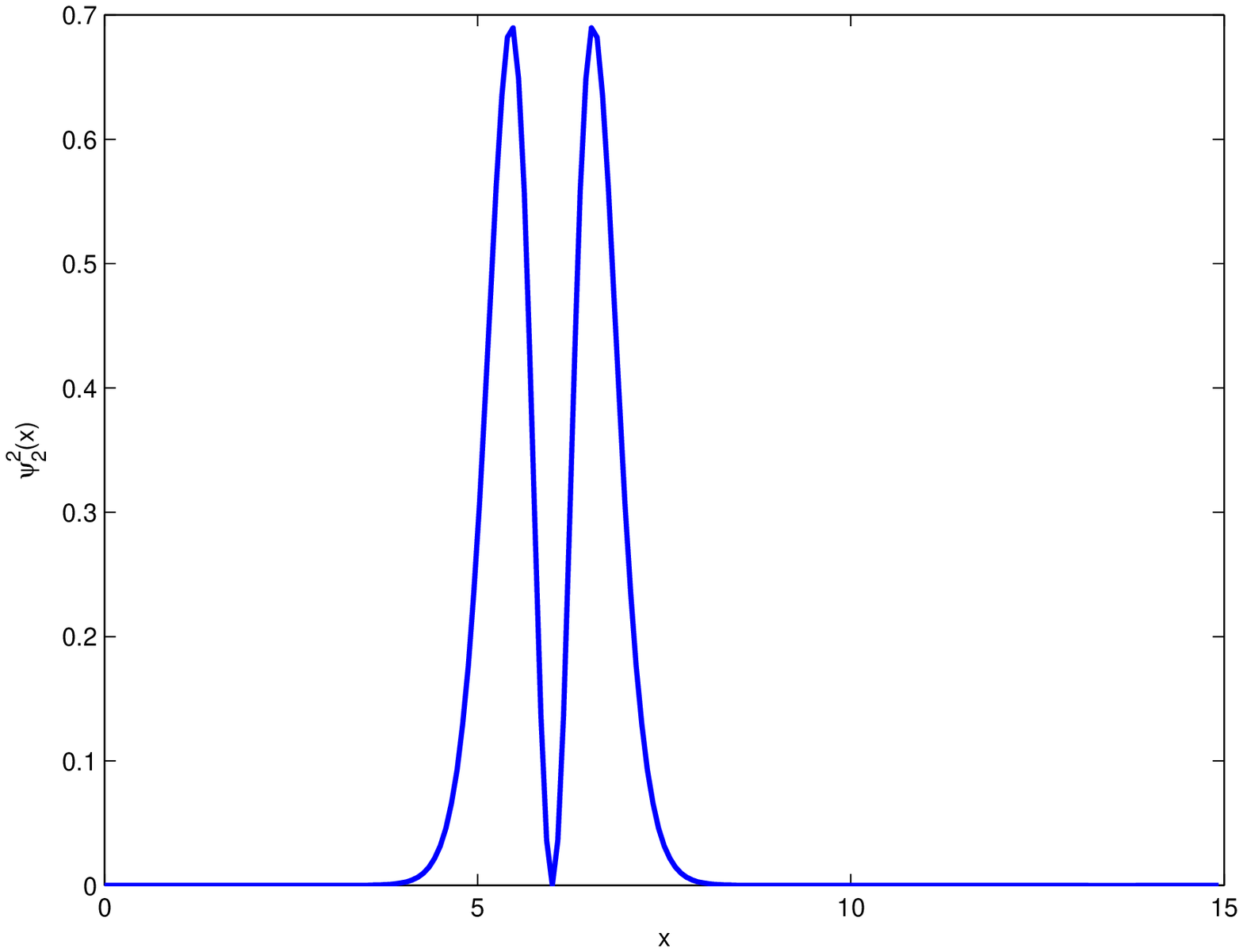,width=5.5cm}}\\
\subfigure {\epsfig{figure=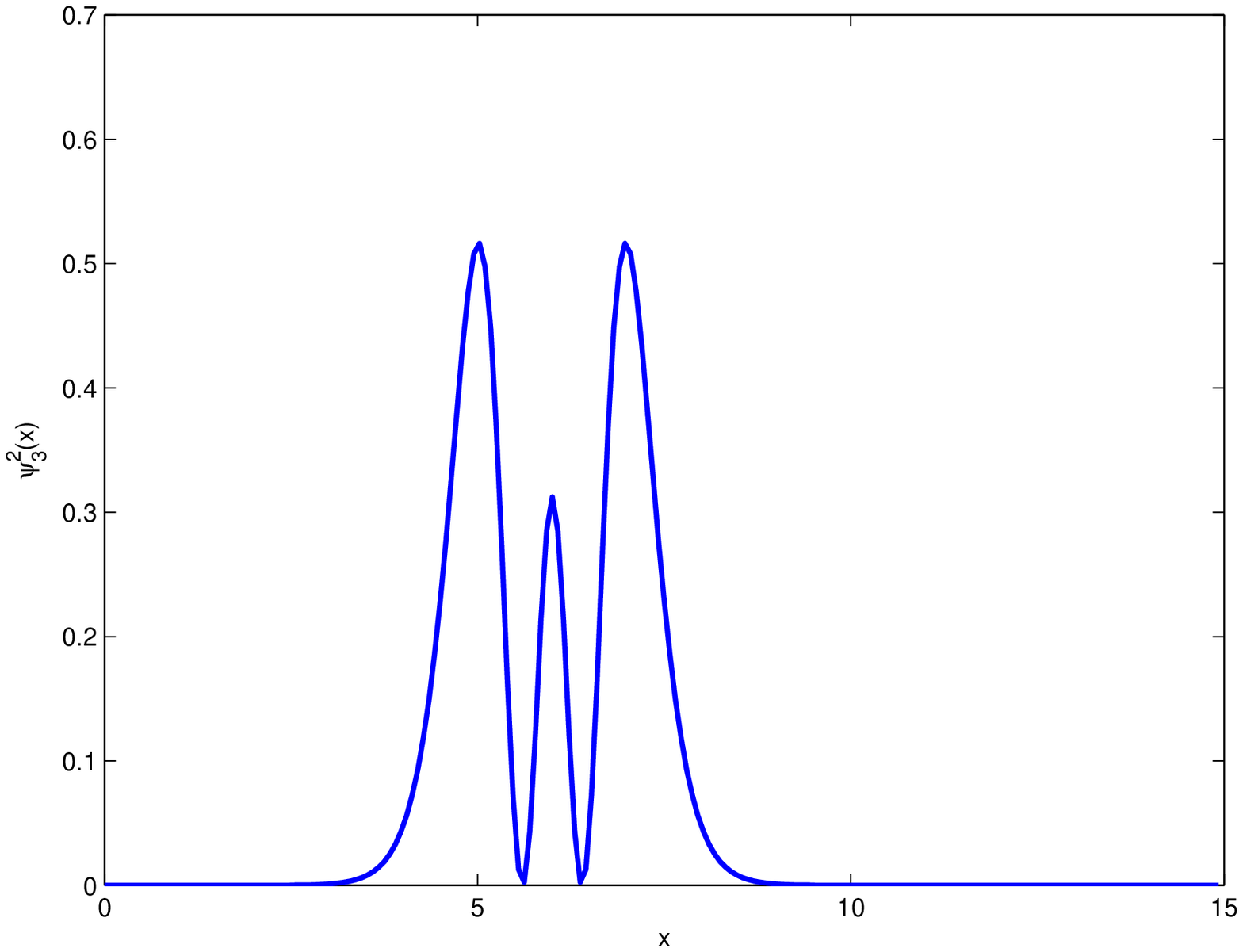,width=5.5cm}}\quad
\subfigure{\epsfig{figure=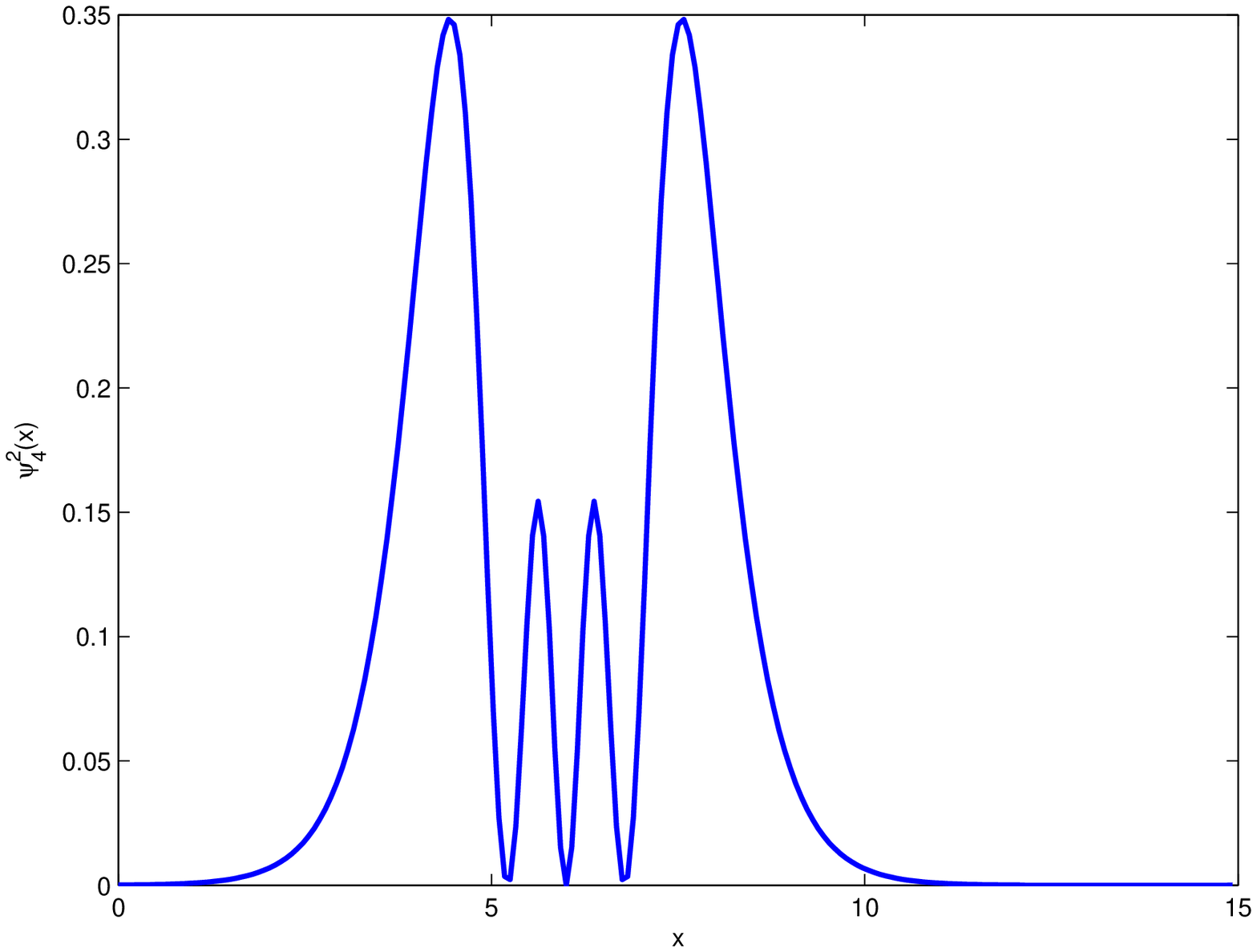,width=5.5cm}} \caption{Four
first eigenfunctions $\psi_{n\chi}^2$ for $\chi=20$ for
$y(x)=\sech^2(x-6)$ in
$\left[0,15\right]$}\label{fonctions_propres}
\end{center}
\end{figure}

Otherwise, in figure \ref{fonctions_propres}, the first
four squared eigenfunctions $\psi_{n\chi}^2$, $n=1,\cdots,4$ are represented for
$\chi=20$. Each $\psi_{n\chi}^2$ has $n-1$
zeros.

\begin{figure}[htbp]
\begin{center}
\subfigure{\epsfig{figure=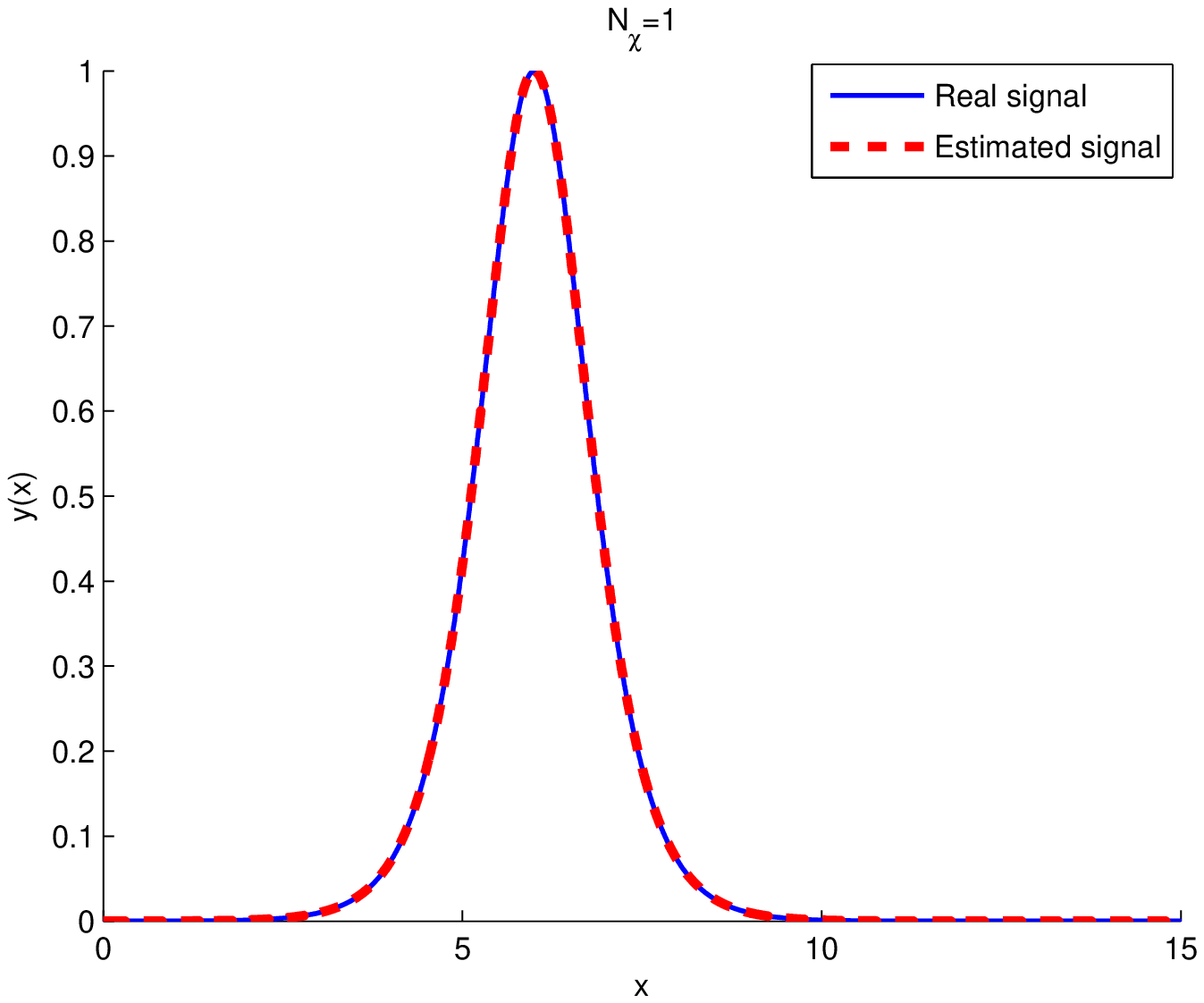,width=5.5cm}}\quad
\subfigure{\epsfig{figure=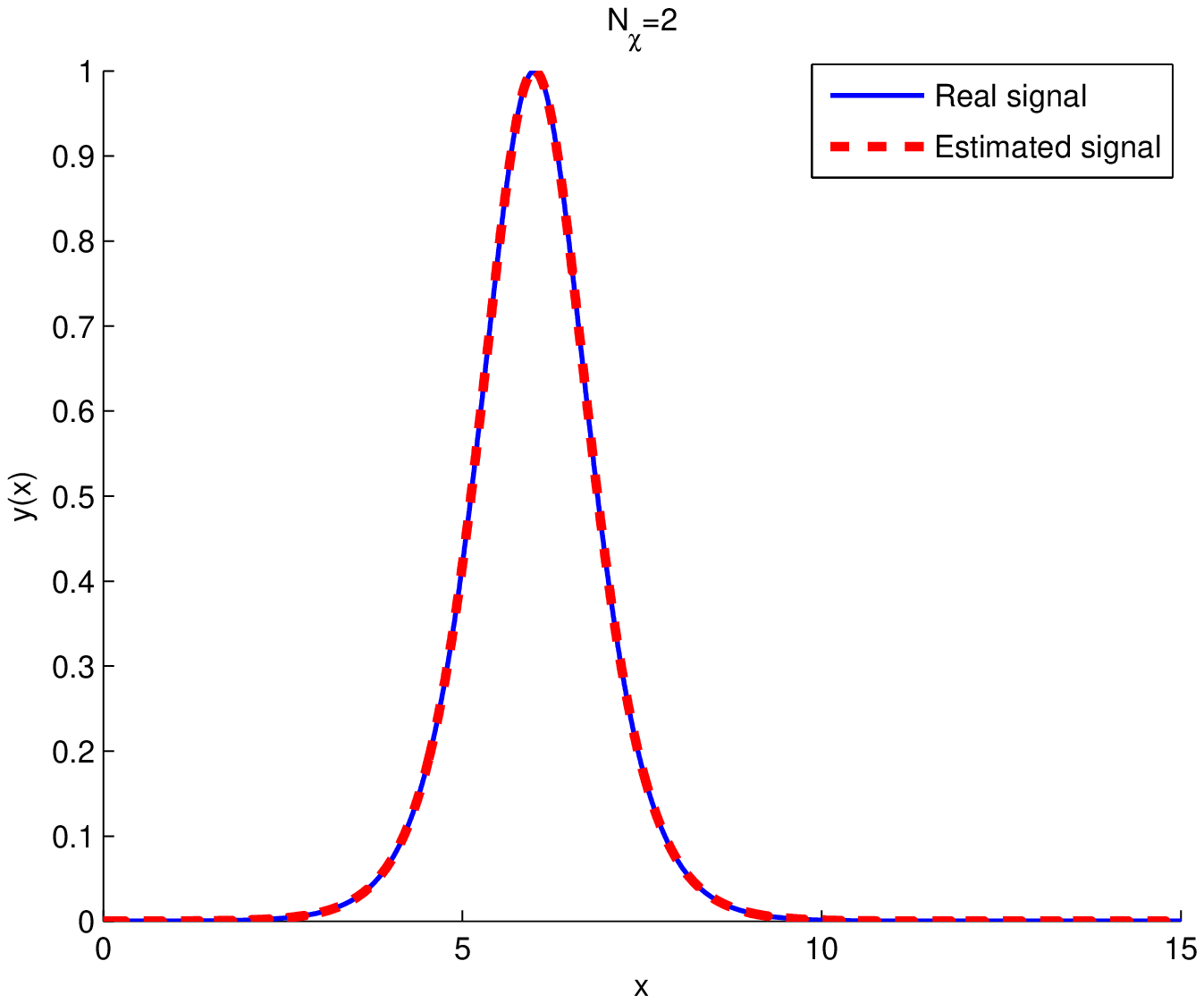,width=5.5cm}}\\
\subfigure {\epsfig{figure=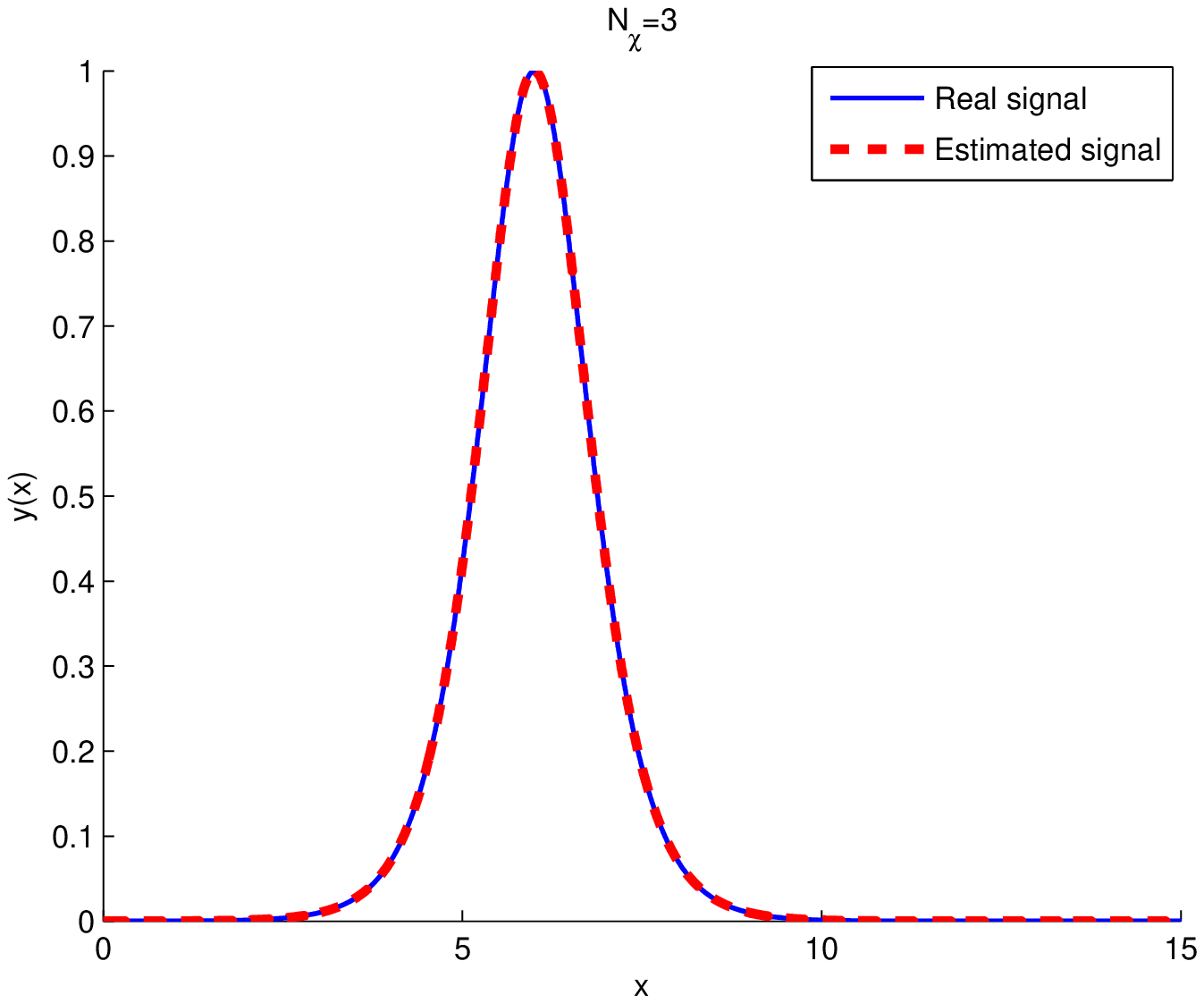,width=5.5cm}}\quad
\subfigure{\epsfig{figure=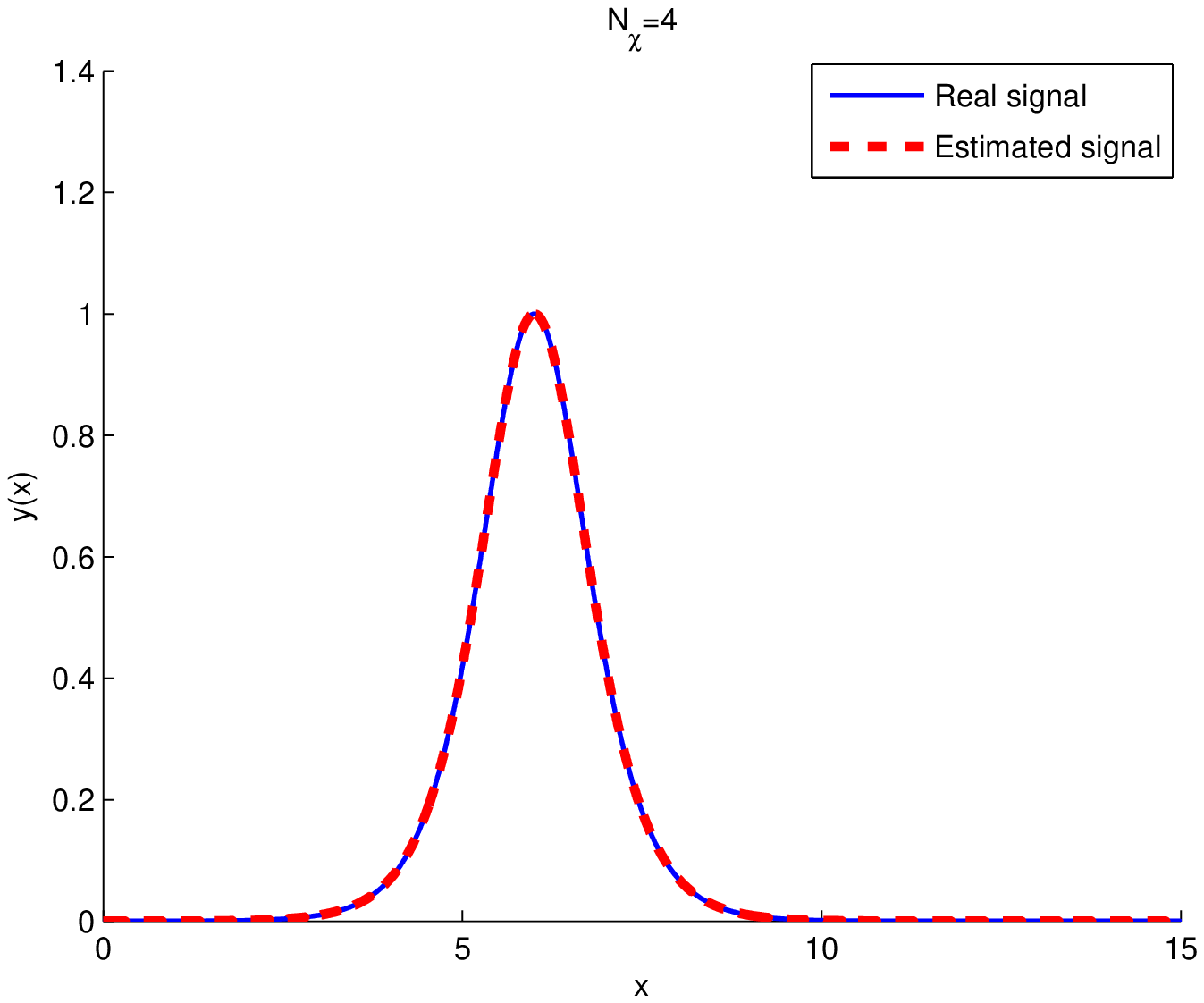,width=5.5cm}}
\caption{Estimation of $y(x)=\sech^2(x-6)$ in $\left[0,15\right]$,
from the left to the right: $N_\chi=1$, $N_\chi=2$ (top).
$N_\chi=3$, $N_\chi=4$ (bottom)}\label{sech2_SINC}
\end{center}
\end{figure}

Figure \ref{sech2_SINC} shows a satisfactory reconstruction of $y$ for $N_\chi= 1, 2, 3$ and $4$.

%\begin{table}
%\begin{center}
%\begin{tabular}{ll}
% \includegraphics[width=7.5cm]{Figures/sech2_erreur_SINC}
% & \includegraphics[width=7cm]{Figures/vap_sech2_SINC}\\
% \hspace{1cm} (a)& \hspace{1cm}(b)  \\
%\end{tabular}
%\end{center}
%\caption{(a) Mean square error and number of negative eigenvalues  according to $\chi$ for $y(x)=\sech^2(x-6)$ in
%$\left[0,15\right]$.  (b) Four first eigenvalues according to  $\chi$ for $y(x)=\sech^2(x-6)$ in $\left[0,15\right]$} \label{erreur_sech2_SINC}
%\end{table}

%
%\begin{figure}[htbp]
%  \begin{center}
%\includegraphics[width=10cm]{Figures/sech2_erreur_SINC}
%\caption{Mean square error and number of negative eigenvalues
%according to $\chi$ for $y(t)=\sech^2(t-6)$ in
%$\left[0,15\right]$}\label{erreur_sech2_SINC}
%\end{center}
%\end{figure}
%
%\begin{figure}[htbp]
%  \begin{center}
%\includegraphics[width=10cm]{Figures/vap_sech2_SINC}
%\caption{Four first eigenvalues according to $\chi$ for
%$y(t)=\sech^2(t-6)$ in $\left[0,15\right]$}\label{vap_sech2_SINC}
%\end{center}
%\end{figure}

\subsection{Estimation of some signals}

In this section, we are interested in the estimation of some
signals with the SCSA. In each case, we represent the estimation
error, the number of negative eigenvalues according to $\chi$ and
the  real and estimated signals for different values of $\chi$.

We start with a gaussian signal given by:
\begin{equation}\label{eq1}
   y(x)=\frac{1}{\sigma\sqrt{2\pi}}e^{-\frac{(x-\mu)^2}{2\sigma^2}}.
\end{equation}
For the numerical tests we take $\sigma=0.1$ and $
\mu=0.75$.

\begin{figure}[htbp]
  \begin{center}
\includegraphics[width=9cm]{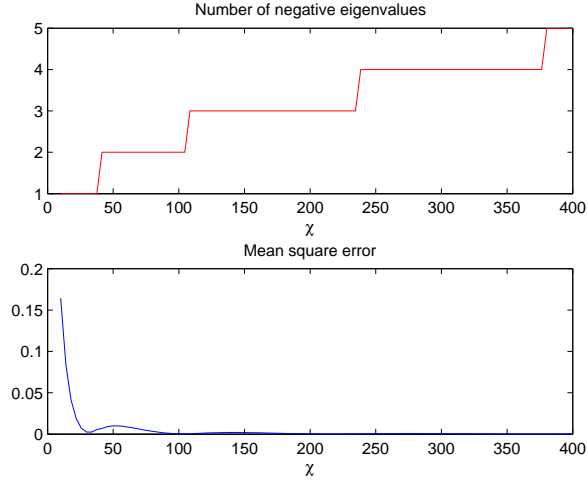}
\caption{Mean square error and number of negative eigenvalues
according to  $\chi$ for a gaussian
signal}\label{erreur_gaussienne}
\end{center}
\end{figure}

\begin{figure}[htbp]
\begin{center}
\subfigure{\epsfig{figure=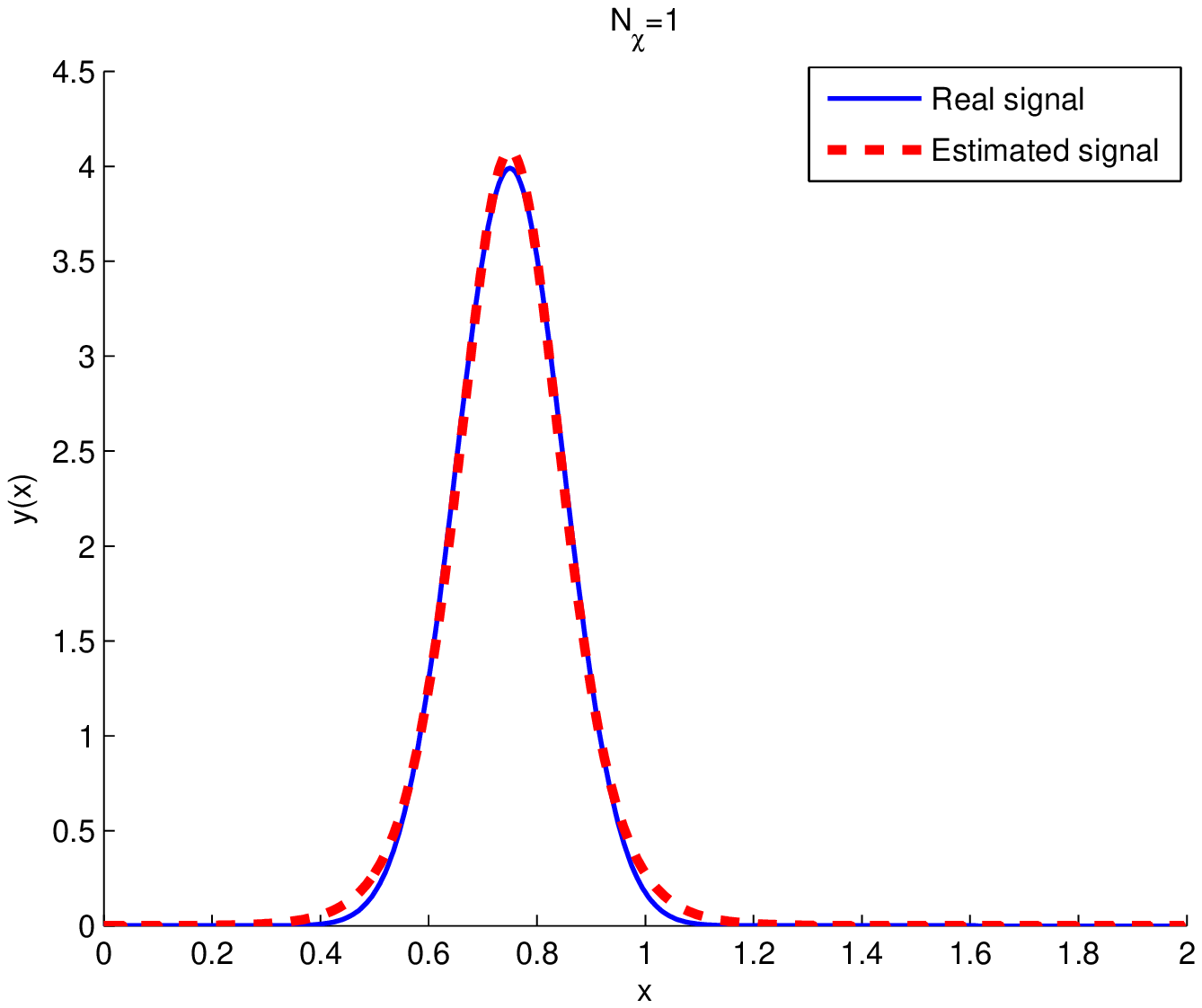,width=5cm}}\quad
\subfigure{\epsfig{figure=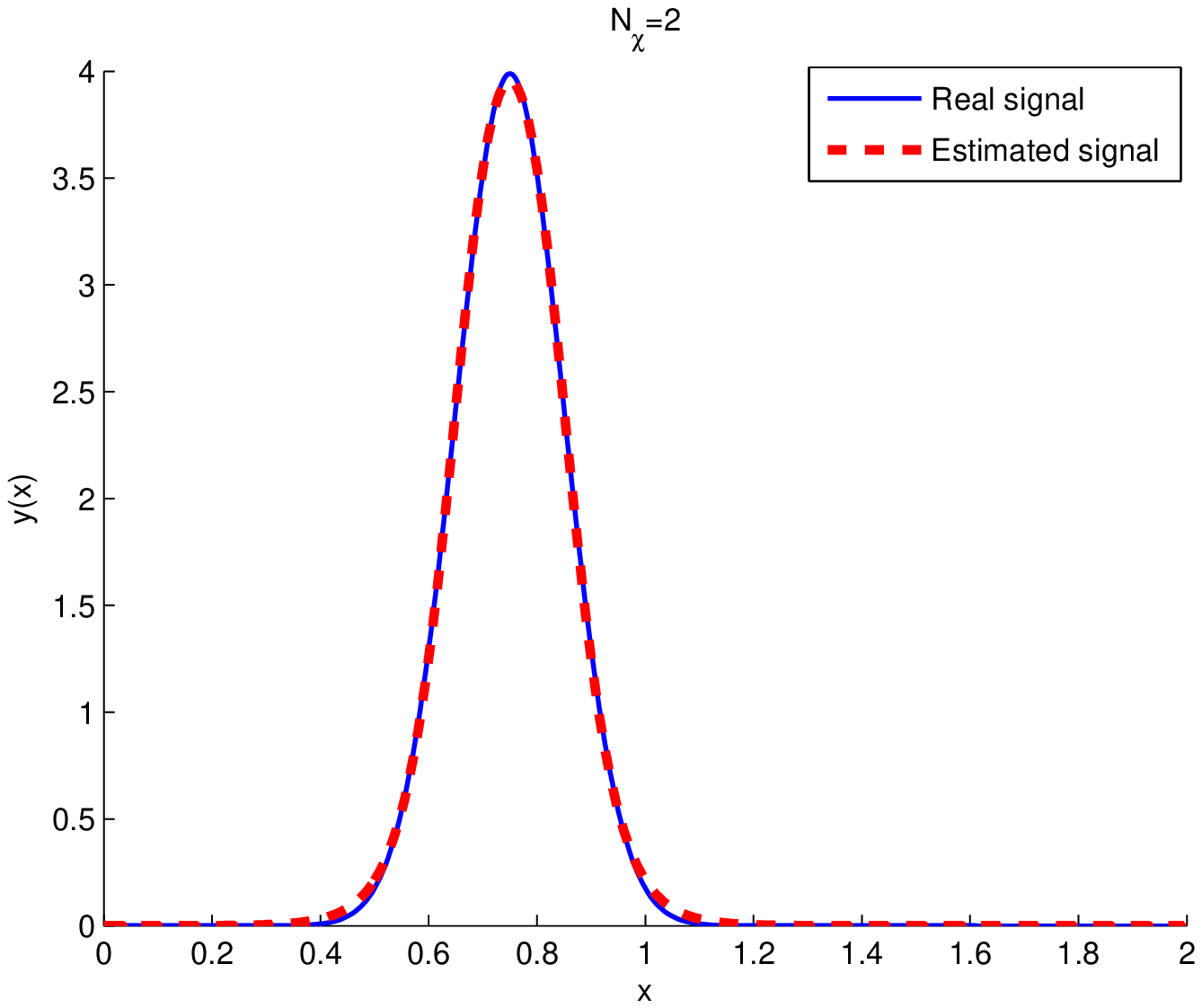,width=5cm}}\\
\subfigure {\epsfig{figure=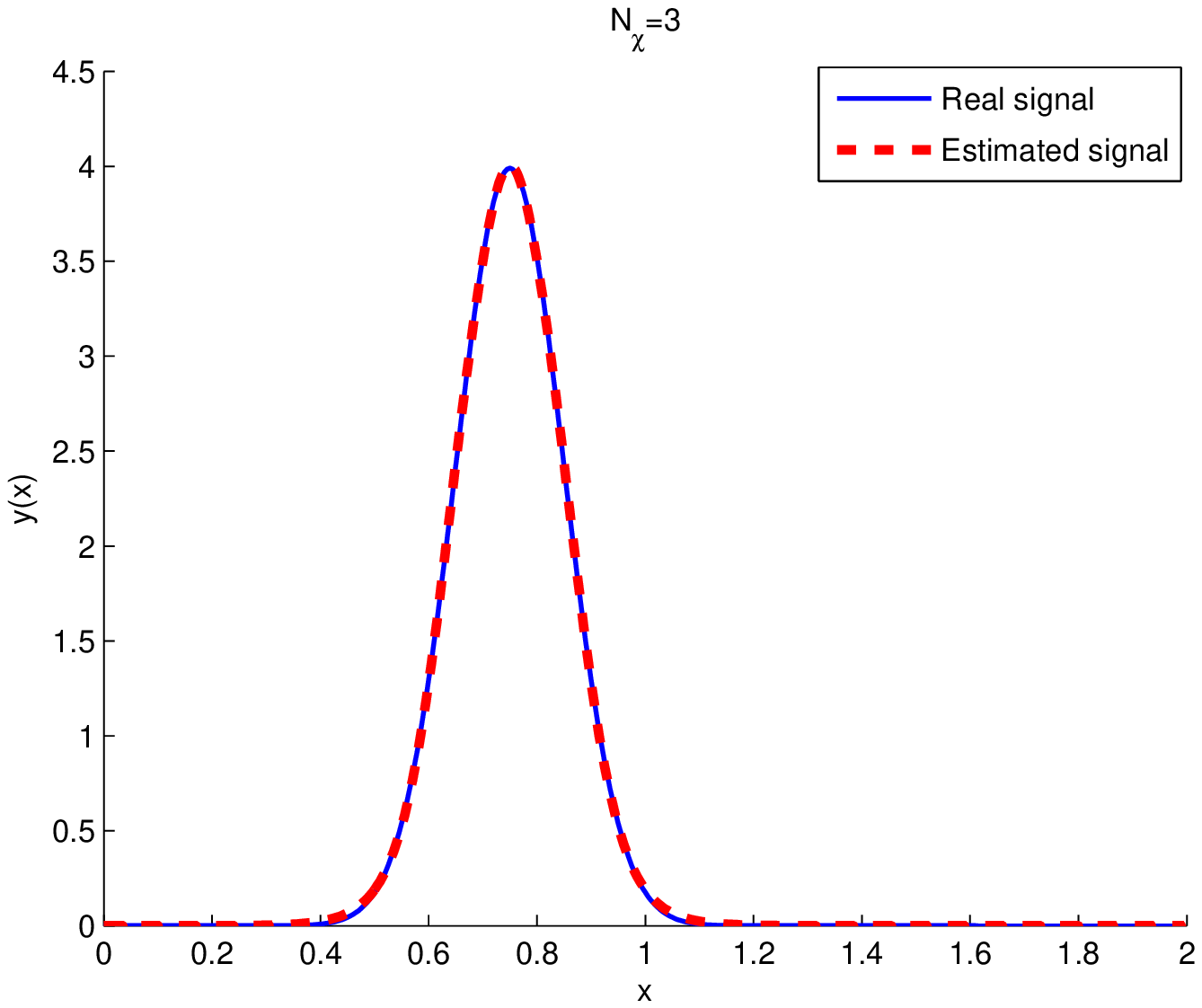,width=5cm}}\quad
\subfigure{\epsfig{figure=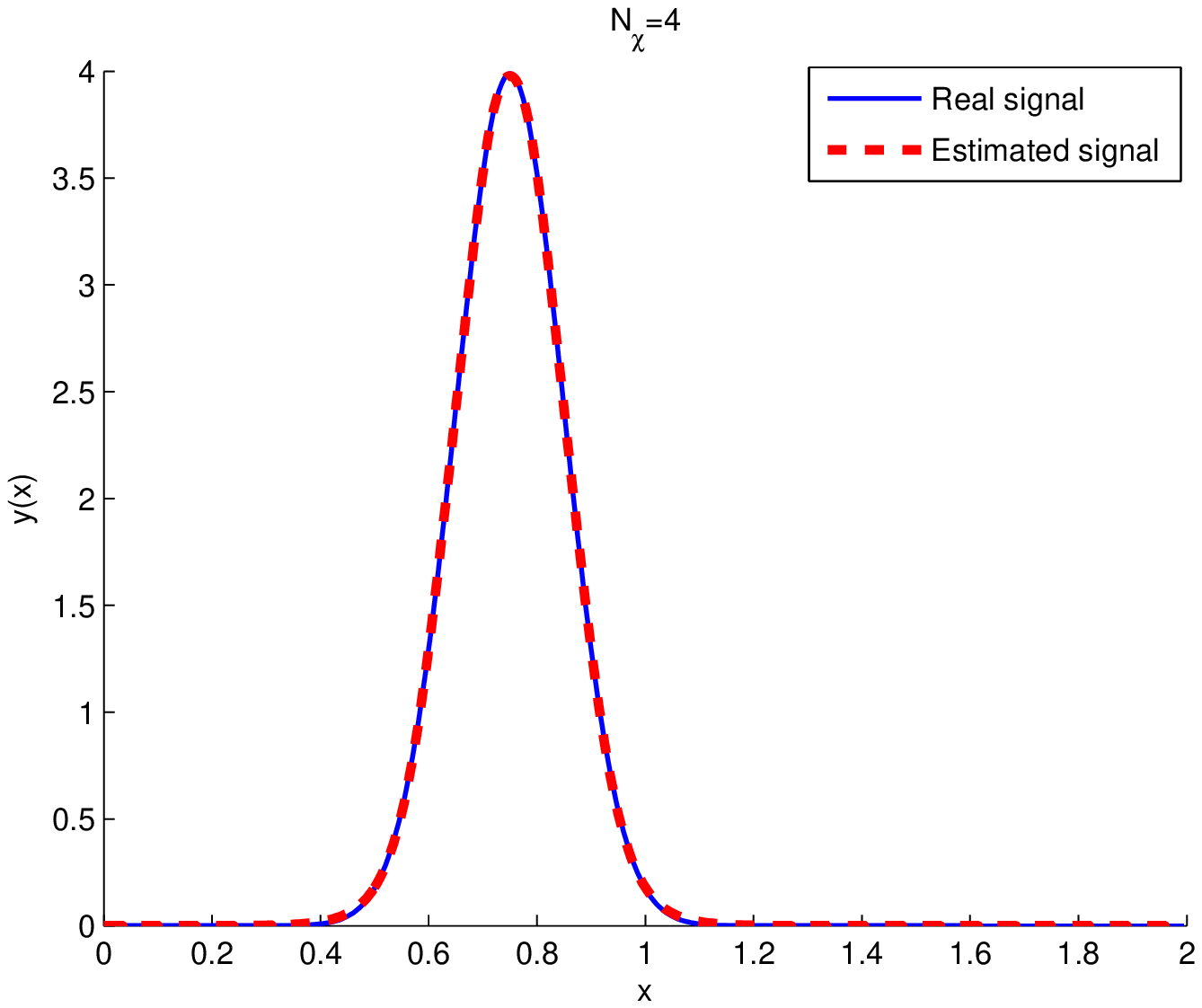,width=5cm}}
\caption{Estimation of a gaussian signal, from the left to the
right: $N_\chi=1$, $N_\chi=2$ (top). $N_\chi=3$, $N_\chi=4$
(bottom)}\label{gaussian}
\end{center}
\end{figure}

Figures \ref{erreur_gaussienne} and \ref{gaussian} illustrate the
results. We notice that with $N_\chi=2$, the estimation
is satisfactory and as $N_\chi$ increases better
is the approximation.

Now we are interested in a sinusoidal signal defined  in a finite
interval $I$,
\begin{equation}\label{sin}
y(x)=\left\{
  \begin{array}{cc}
    A \sin(\omega x+\phi) & x\in I \\
    0 & \mbox{otherwise} \\
  \end{array}
\right.
\end{equation}
This signal has negative values, so to apply the SCSA, we must
translate the signal by $y_{min}=-A$ such that $y-y_{min}
> 0$. The Schr\"{o}dinger operator potential to be considered is then
given by $-\chi(y-y_{min})$. For the numerical tests, we took $A=2$, $\omega=\pi$
and $\phi=-0.5$.

\begin{figure}[htbp]
  \begin{center}
\includegraphics[width=9cm]{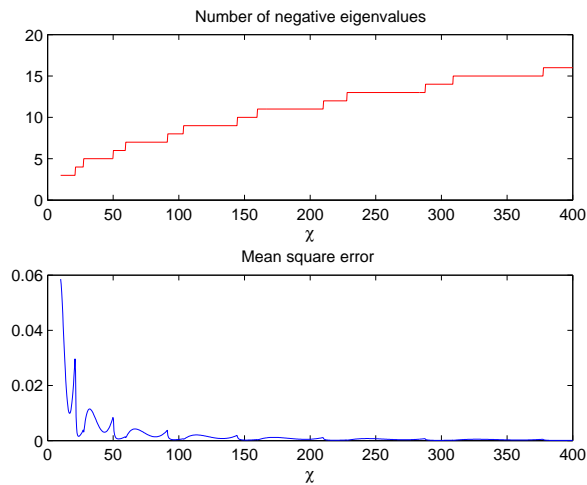}
\caption{Mean square error and number of negative eigenvalues
according to $\chi$ for a  sinusoidal signal}\label{error_sinus}
\end{center}
\end{figure}

\begin{figure}[htbp]
\begin{center}
\subfigure{\epsfig{figure=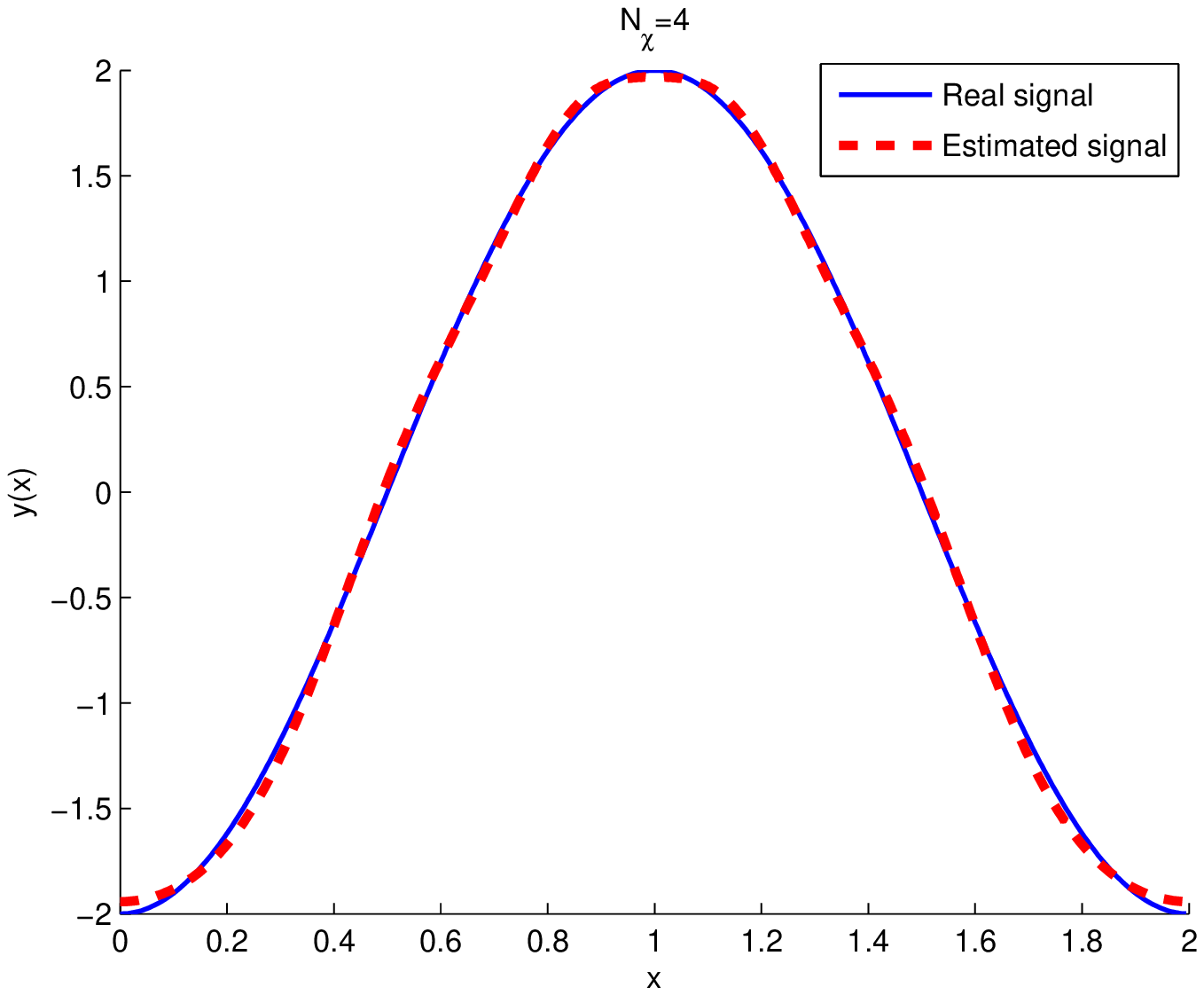,width=5cm}}\quad
\subfigure{\epsfig{figure=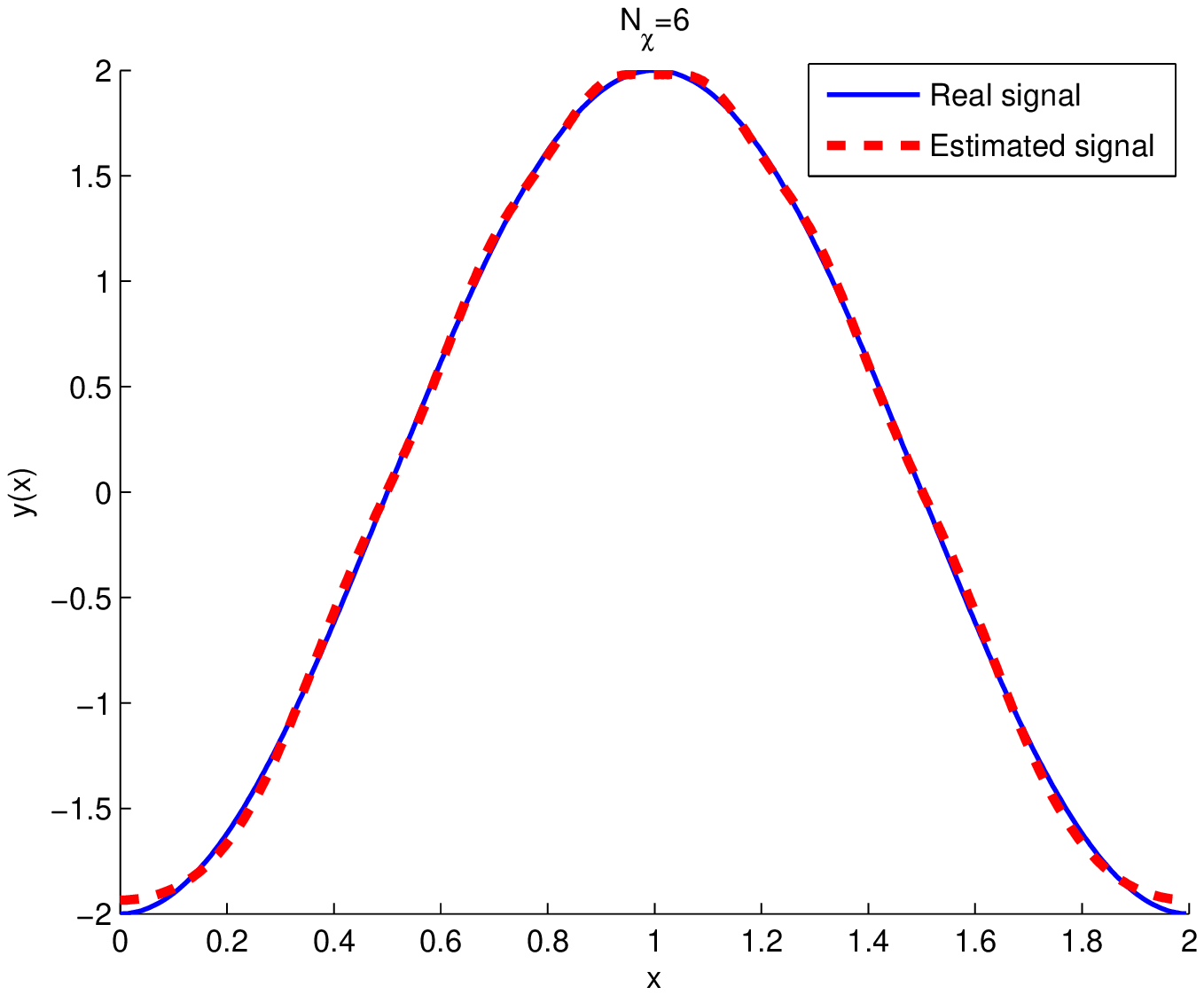,width=5cm}}\\
\subfigure {\epsfig{figure=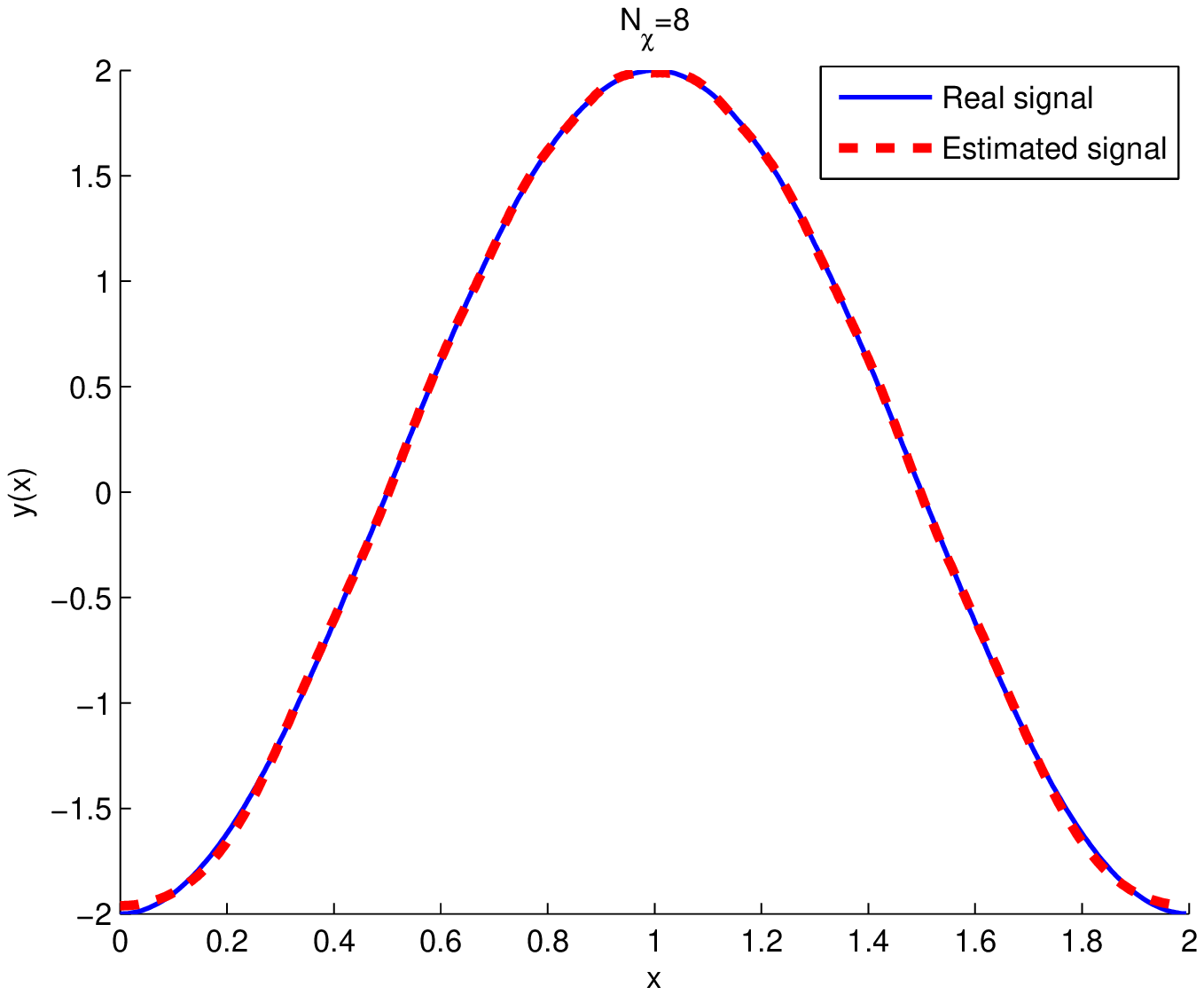,width=5cm}}\quad
\subfigure{\epsfig{figure=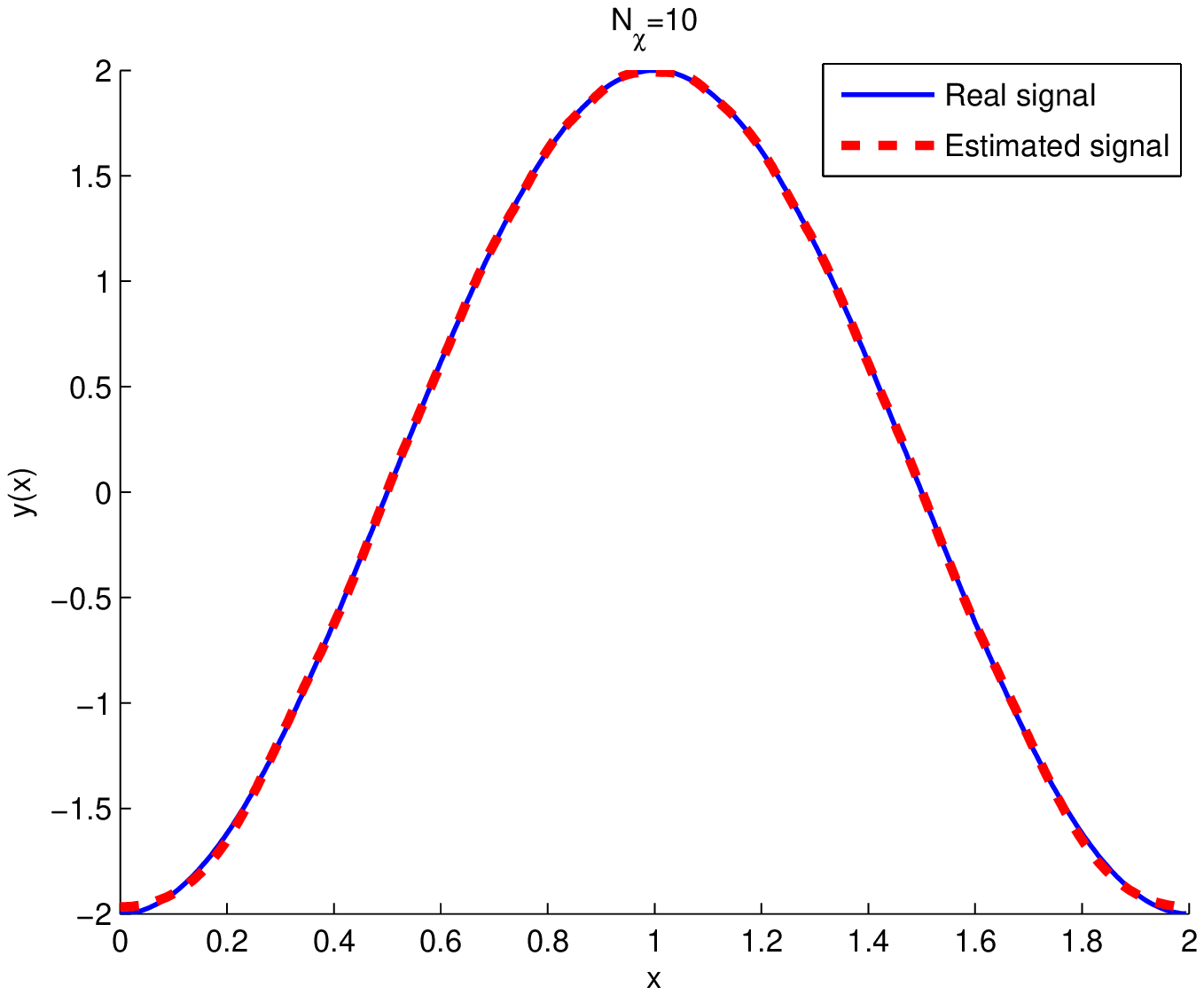,width=5cm}}
\caption{Estimation of a sinusoidal signal, from the left to the
right: $N_\chi=4$, $N_\chi=6$ (top). $N_\chi=8$, $N_\chi=10$
(bottom)}\label{sinus}
\end{center}
\end{figure}

\begin{figure}[htbp]
\begin{center}
\includegraphics[width=7cm]{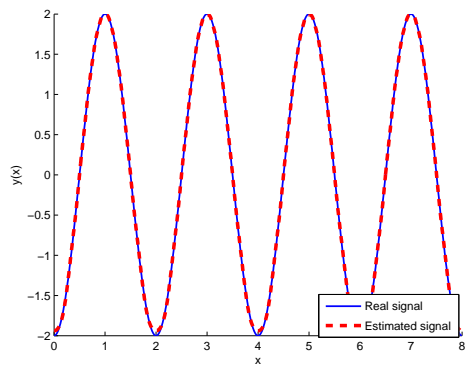}\\
\caption{Estimation of 4 sinusoidal signal periods with
$N_\chi=10$}\label{sin_plu}\end{center}
\end{figure}

The results are represented in figures \ref{error_sinus},
\ref{sinus} and  \ref{sin_plu}. In \ref{sinus}, a single period of
the signal is considered while in \ref{sin_plu}, four periods are
represented. In the last case we noticed that the negative
eigenvalues are of multiplicity 4, they are repeated in each
period.

\begin{figure}[htbp]
  \begin{center}
\includegraphics[width=9cm]{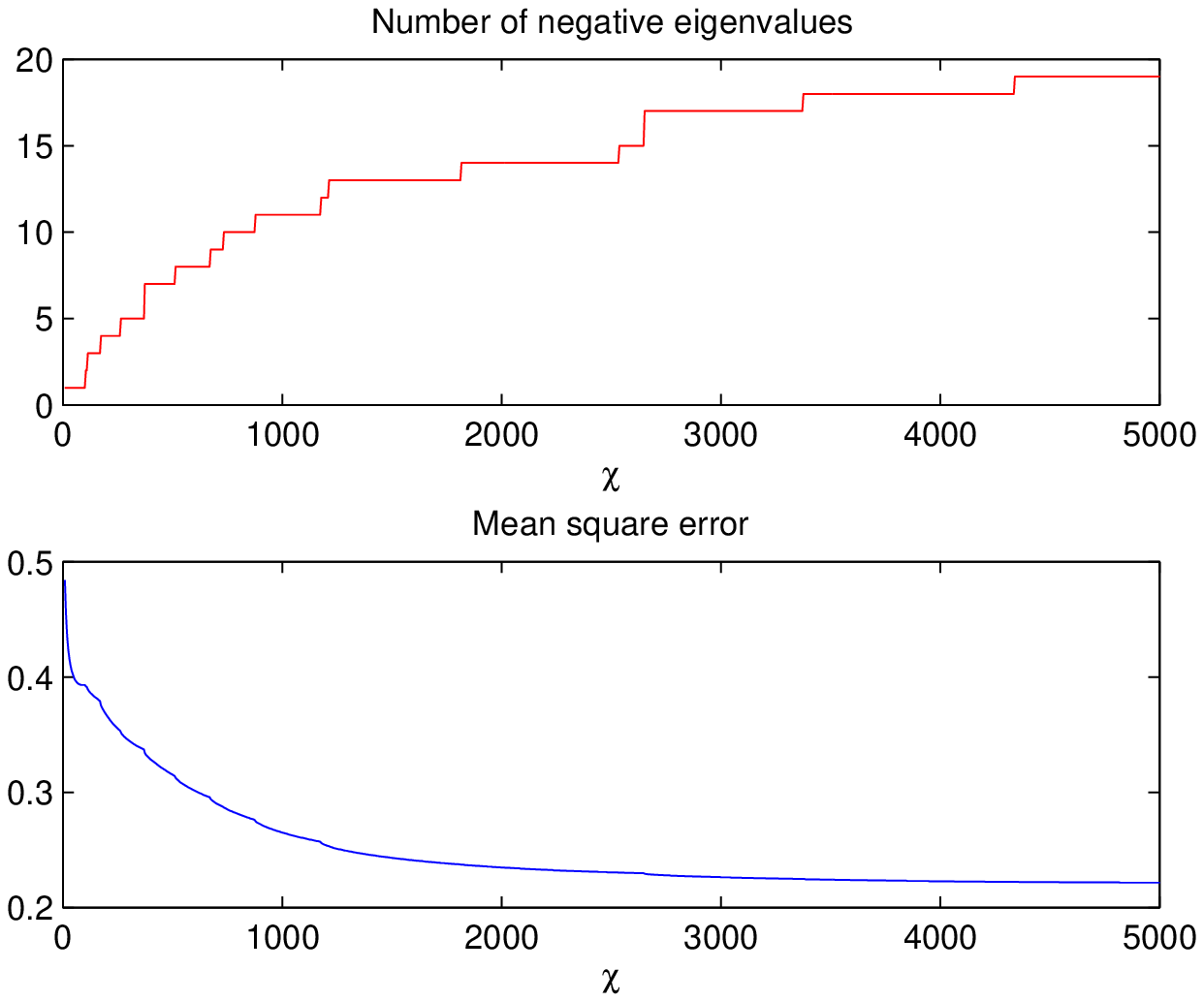}
\caption{Mean square error and number of negative eigenvalues
according to $\chi$ for a chirp signal}\label{error_chirp}
\end{center}
\end{figure}

\begin{figure}[htbp]
\begin{center}
\subfigure{\epsfig{figure=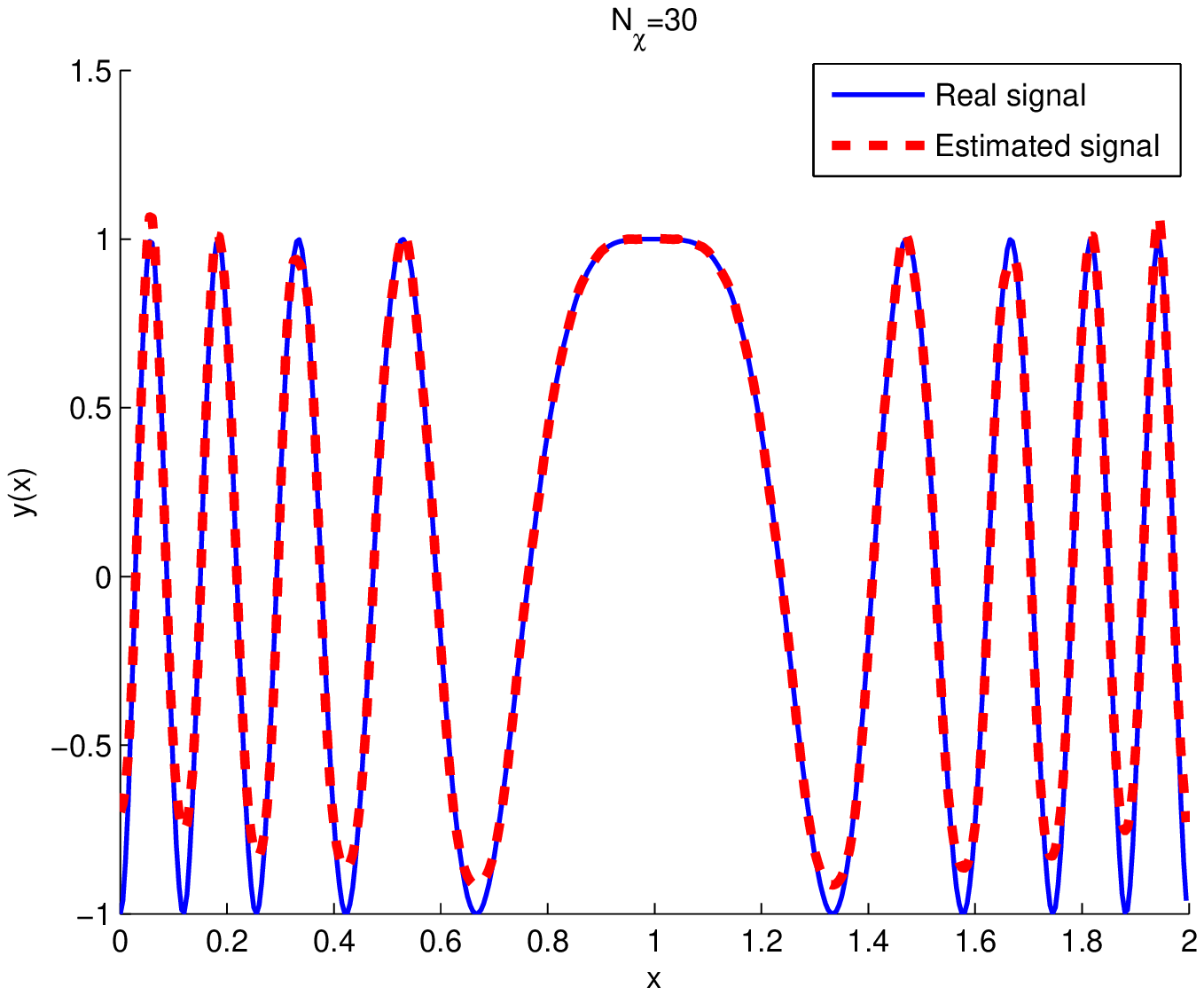,width=5cm}}\quad
\subfigure{\epsfig{figure=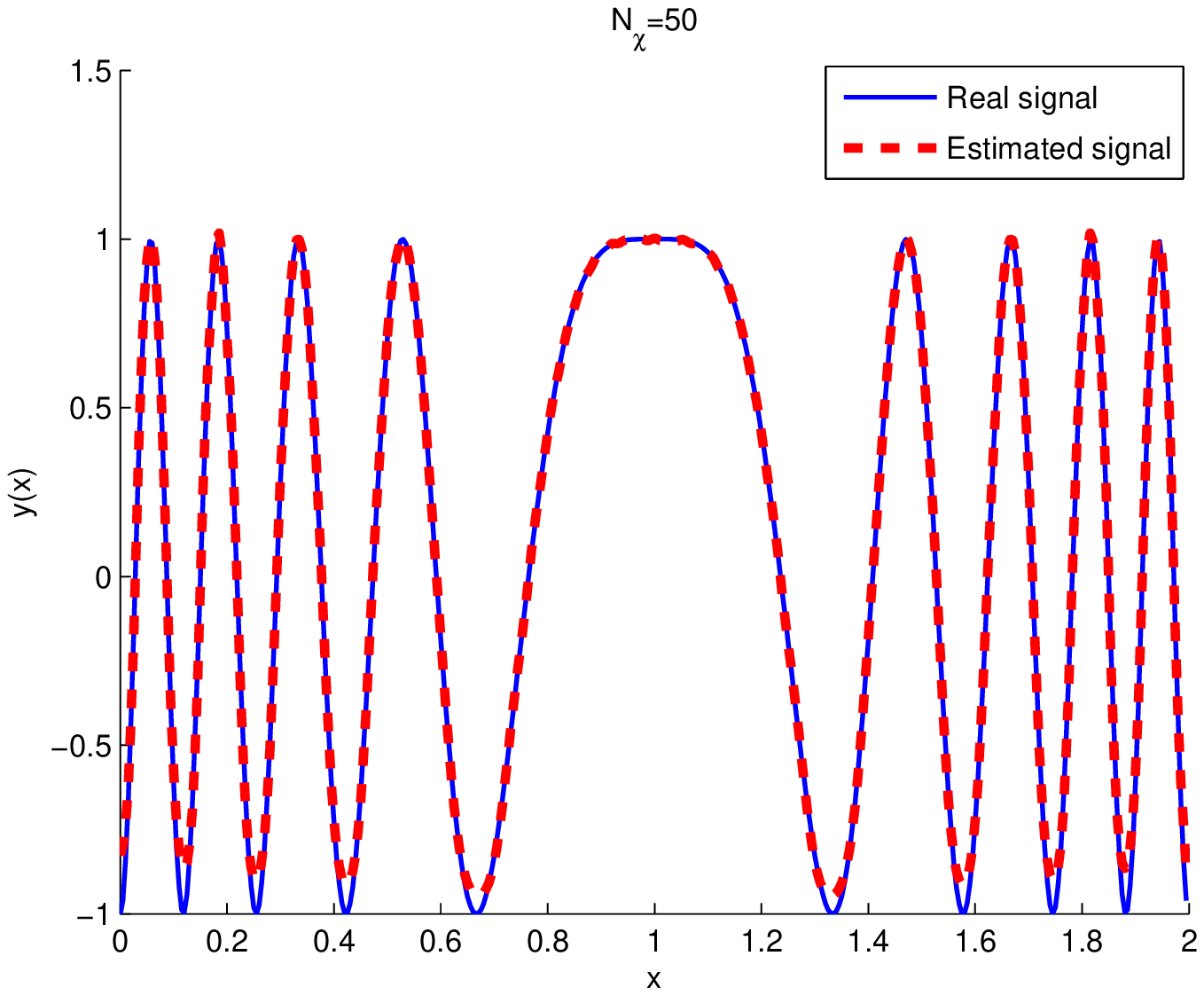,width=5cm}}\\
\subfigure {\epsfig{figure=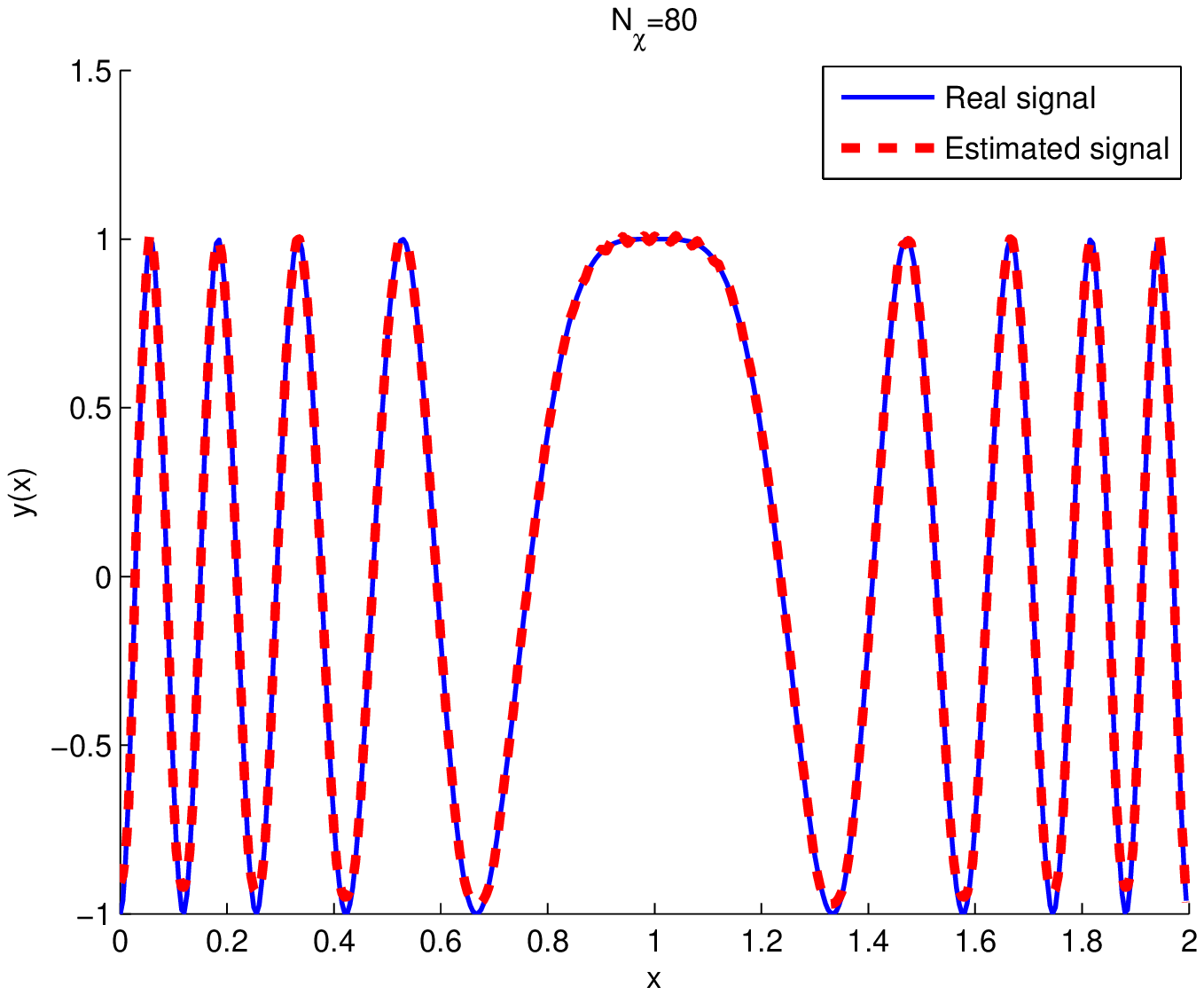,width=5cm}}\quad
\subfigure{\epsfig{figure=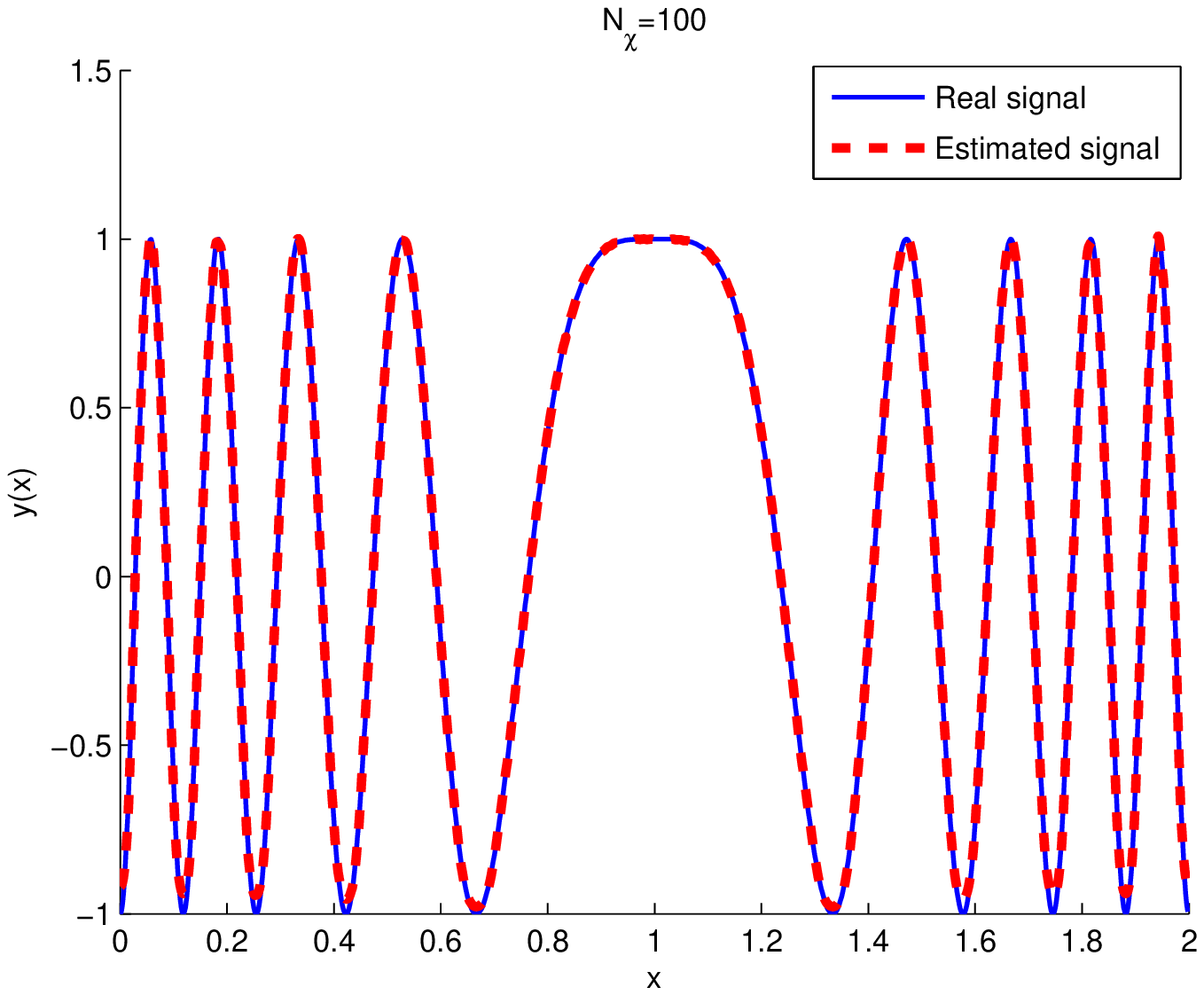,width=5cm}}
\caption{Estimation of a chirp signal, from the left to the right:
$N_\chi=30$, $N_\chi=50$ (top). $N_\chi=80$, $N_\chi=100$
(bottom)}\label{chirp}
\end{center}
\end{figure}

Finally, figures \ref{error_chirp} and \ref{chirp} illustrate the
results obtained in the case of a chirp signal. We recall that a chirp signal
is usually defined by a time varying frequency sinusoid. In our
tests, we considered a linear variation of the frequency.

%
%\begin{figure}[htbp]
%  \begin{center}
%\includegraphics[width=7cm]{Figures/vap_chirp}
%\caption{Evolution des quatre premières valeurs propres en
%fonction de $\chi$ pour un signal chirp}\label{vap_chirp}
%\end{center}
%\end{figure}

%---------------------------------------------------------------------------------------------

\section{Arterial blood pressure analysis with the SCSA}

ABP plays an important role in the cardiovascular system. So
many studies were done aiming at proposing mathematical models in order
to understand the cardiovascular system both in healthy and pathological cases.
Despite the large number of ABP models, the interpretation of ABP in clinical practice
is often restricted to the interpretation of the maximal and the minimal values
called respectively the systolic pressure and the diastolic pressure.
None information on the instantaneous variability of the pressure is given in this case. However,
pertinent information can be extracted from ABP waveform. The SCSA seems to provide a new tool for the analysis of ABP waveform. This section presents some obtained results.

We denote the ABP signal $P$ and $\hat{P}$ its estimation using the SCSA such that

%\begin{equation}\label{pestime}
%    \hat{P}(t)=h\sum_{n=1}^{N_h}{\kappa_{nh}\psi_{nh}^2(t)},
%\end{equation}
%where $-\kappa_{nh}^2$, $n=1,\cdots,N_h$ are the $N_h$
%negative eigenvalues of  $H(h;P)$
%and  $\psi_{nh}$ the associate $L^2-$normalized
%eigenfunctions.

\begin{equation}\label{pestime}
    \hat{P}(t)= \frac{4}{\chi} \sum_{n=1}^{N_\chi}{\kappa_{n\chi}\psi_{n\chi}^2(t)},
\end{equation}
where $-\kappa_{n\chi}^2$, $n=1,\cdots,N_\chi$ are the $N_\chi$
negative eigenvalues of  $H_1(\chi P)$
and  $\psi_{n\chi}$ the associated $L^2-$normalized
eigenfunctions. As ABP signal is a function of time, we use the time variable $t$ in the
Schr\"{o}dinger equation instead of the space variable $x$.

The ABP signal was estimated for several values of the parameter $\chi$ and hence $N_\chi$.  Figure  \ref{pression
doigt2} illustrates the measured and estimated pressures for one beat of ABP and the estimated error with $N_\chi=9$.
Signals measured at the aorta level and the finger respectively were considered.  We point out that $5$ to $9$
negative eigenvalues are sufficient for a good estimation of ABP signals \cite{LaCrPaSo:07}, \cite{LaCrSo:07}.\\

\begin{figure}[htbp]
\begin{center}
\subfigure{\epsfig{figure=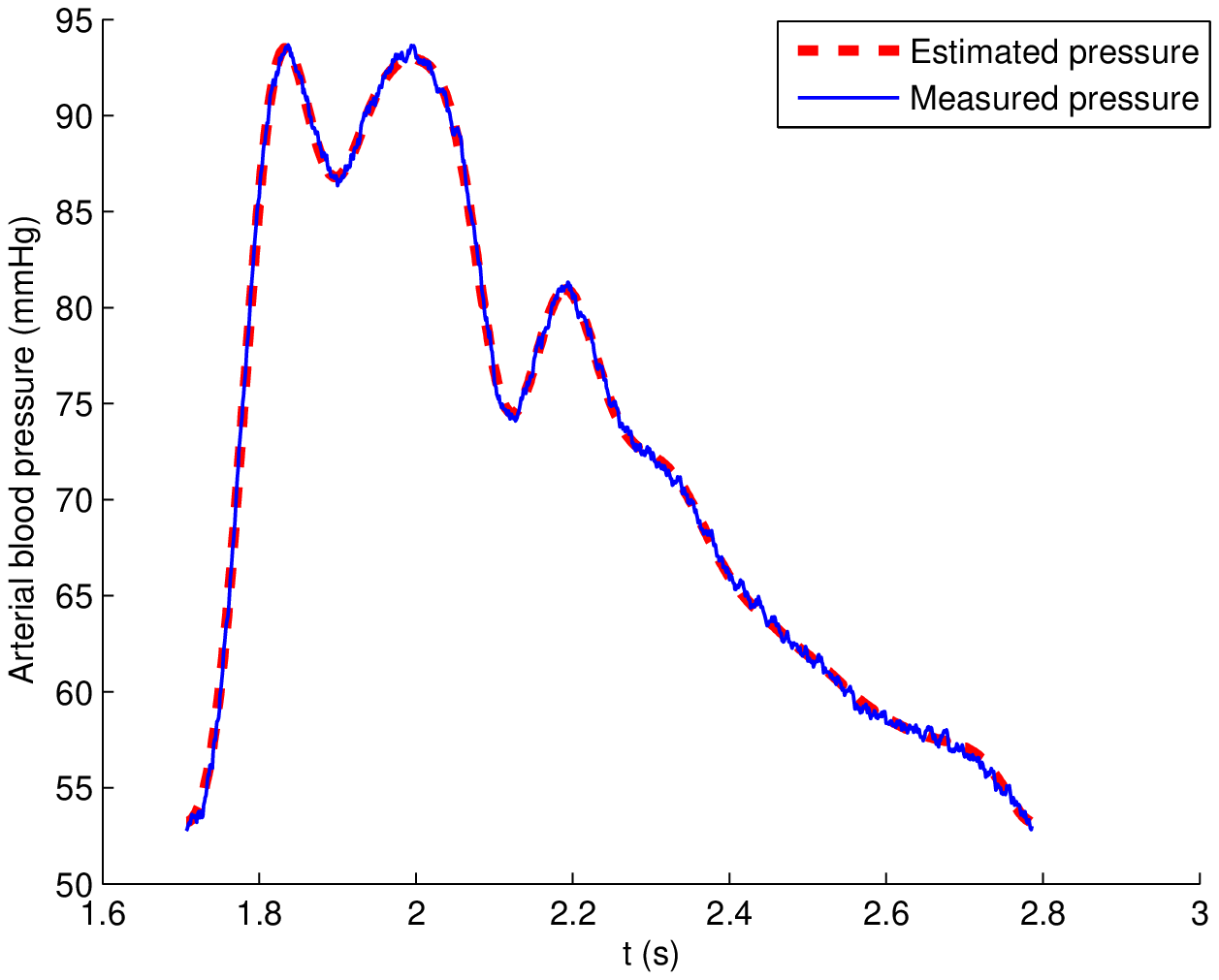,width=5cm}}
\subfigure{\epsfig{figure=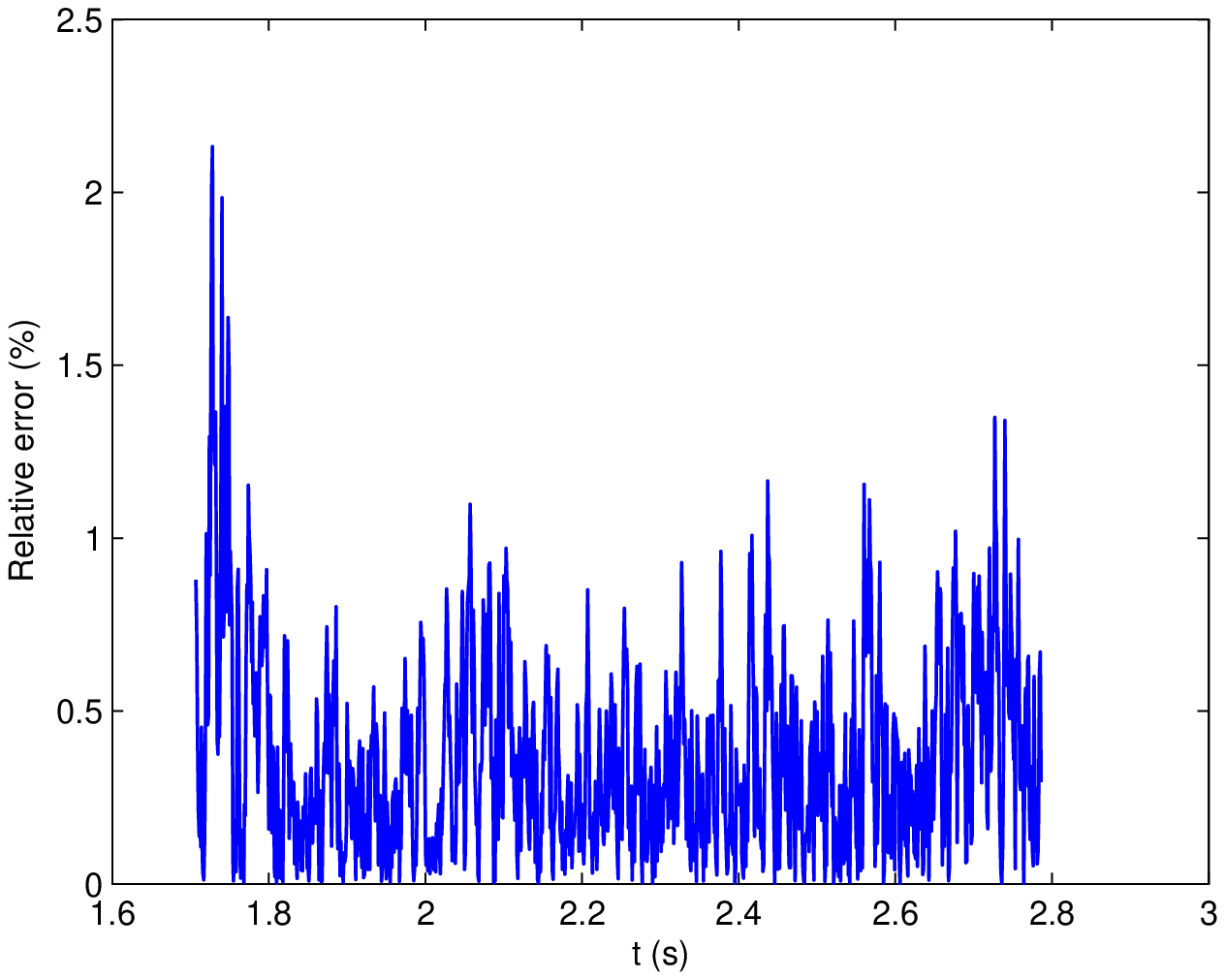,width=5cm}}
\begin{center}
{(a) Aorta}\end{center}
\subfigure{\epsfig{figure=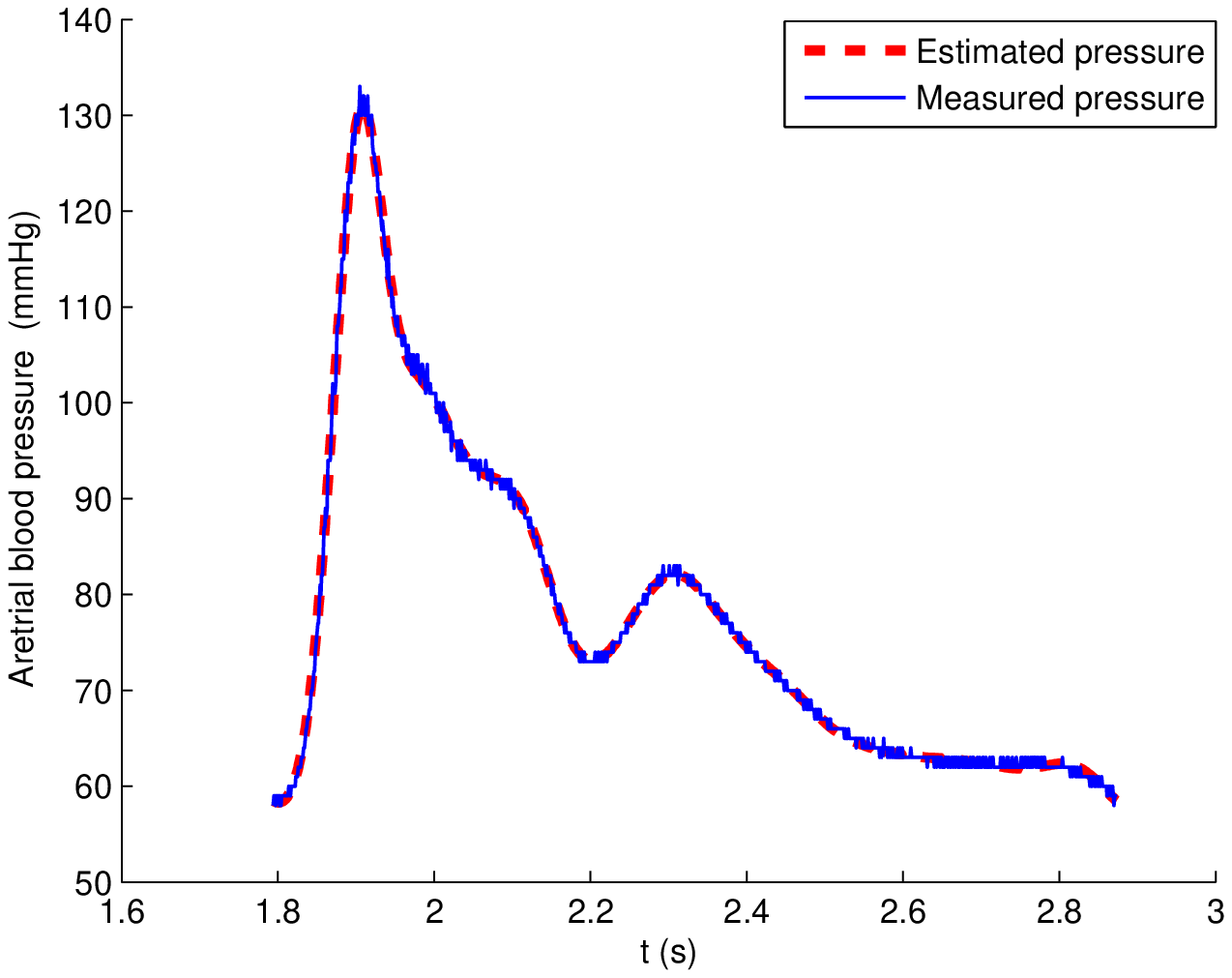,width=5cm}}
\subfigure{\epsfig{figure=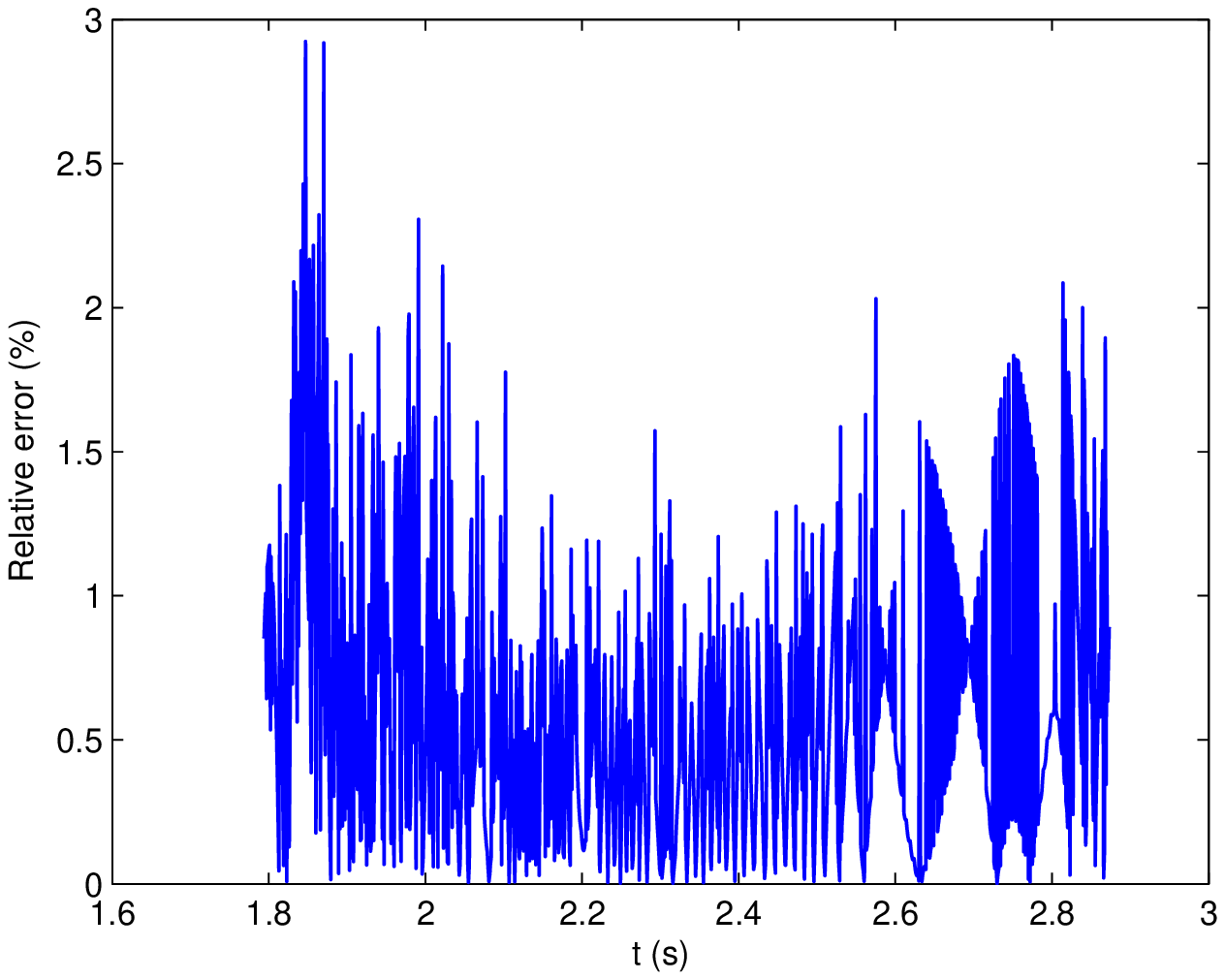,width=5cm}}
\begin{center}
{(b) Finger}\end{center}
 \caption{Reconstruction of the pressure at the aorta and the finger level with the SCSA and $N_\chi=9$. On the left the estimated and measured pressures, On the right the relative error}\label{pression doigt2}
\end{center}
\end{figure}

 A first interest in using the SCSA for  ABP analysis is to decompose the signal into its systolic and diastolic parts respectively. This application was inspired from a reduced model of ABP based on solitons introduced in  \cite{CrSo:07}, \cite{LaCrSo:07J}. Solitons are in fact solutions of some nonlinear partial derivative equations for instance the Korteweg-de Vries (KdV) equation which was considered in this reduced model \cite{CrSo:07}. This model proposes to write the ABP as the sum of two terms: and N-soliton, solution of the KdV equation describing fast phenomena that predominate during the systolic phase
and a two-elements windkessel model that describes slow phenomena during the diastolic phase. Moreover the KdV equation can be solved with the Inverse Scattering Transform (IST) whose definition is recalled in appendix A. In this approach, the KdV equation is associated to a one dimensional Schr\"{o}dinger potential parameterized by time where the potential is given by the solution of the KdV equation at a given time. Therefore a relation between the Schr\"{o}dinger operator and solitons was found \cite{GaGrKrMi:74}: solitons are reflectionless potentials.
 Then according to proposition \ref{approximation exacte}, the SCSA coincides with a soliton representation of a signal for a finite $\chi$ when $- \chi y $ is an $N_\chi$-soliton, where $N_\chi$ denotes the number of negative eigenvalues \footnote{Each soliton is characterized by a negative eigenvalue} of $H_1(\chi y)$.  So each spectral component represents a single soliton.
We know that solitons are characterized by their velocity which is determined by the negative eigenvalues
$-\kappa_{n\chi}^2$, $n=1,\cdots, N_\chi$ of the Schr\"{o}dinger operator. The largest values  $\kappa_{n\chi}$
characterize fast components and the small values of  $\kappa_{n\chi}$ characterize slow components.
From these remarks, we propose to decompose equation (\ref{formule chi}) into two partial sums: the first one, composed of the  $N_s$ ($N_s=1, 2, 3$ in general) largest  $\kappa_{n\chi}$
and the second partial sum composed of the remaining components. The first partial sum describes the systolic phase and the second
one describes the diastolic phase. We denote $\hat{P}_s$ and
$\hat{P}_d$ the systolic pressure and the diastolic pressure respectively estimated with the SCSA. Then we have
\begin{equation}
    \hat{P}_s(t) = \frac{4}{\chi}\sum_{n=1}^{N_{s}}{\kappa_{n\chi}\psi_{n\chi}^2(t)},
    \quad \quad \quad \hat{P}_d(t) = \frac{4}{\chi} \sum_{n=N_{s}+1}^{N_\chi}{\kappa_{n\chi}\psi_{n\chi}^2(t)}.
\end{equation}

\begin{figure}[htbp]
\begin{center}
\subfigure{\epsfig{figure=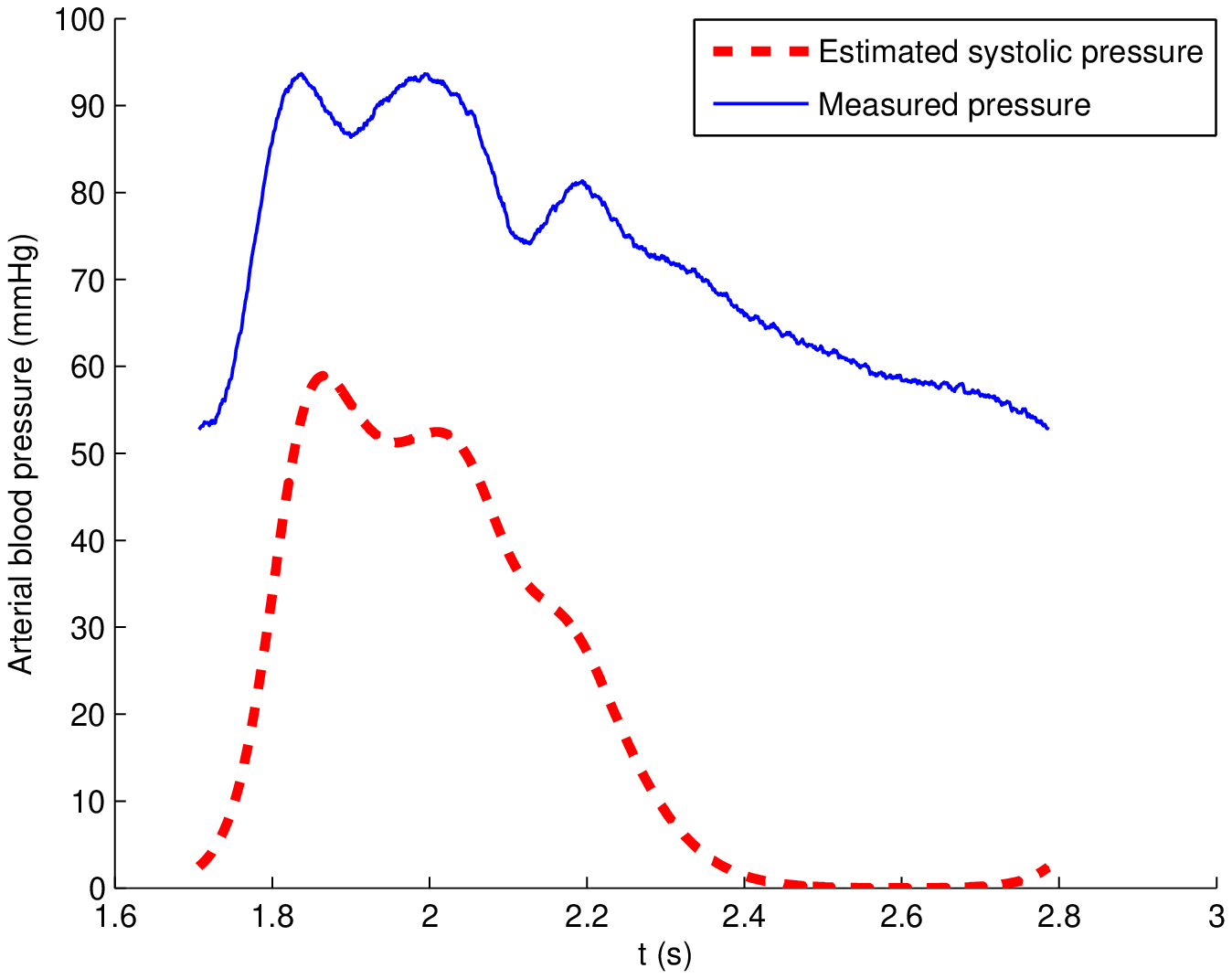,width=6cm}}
\subfigure{\epsfig{figure=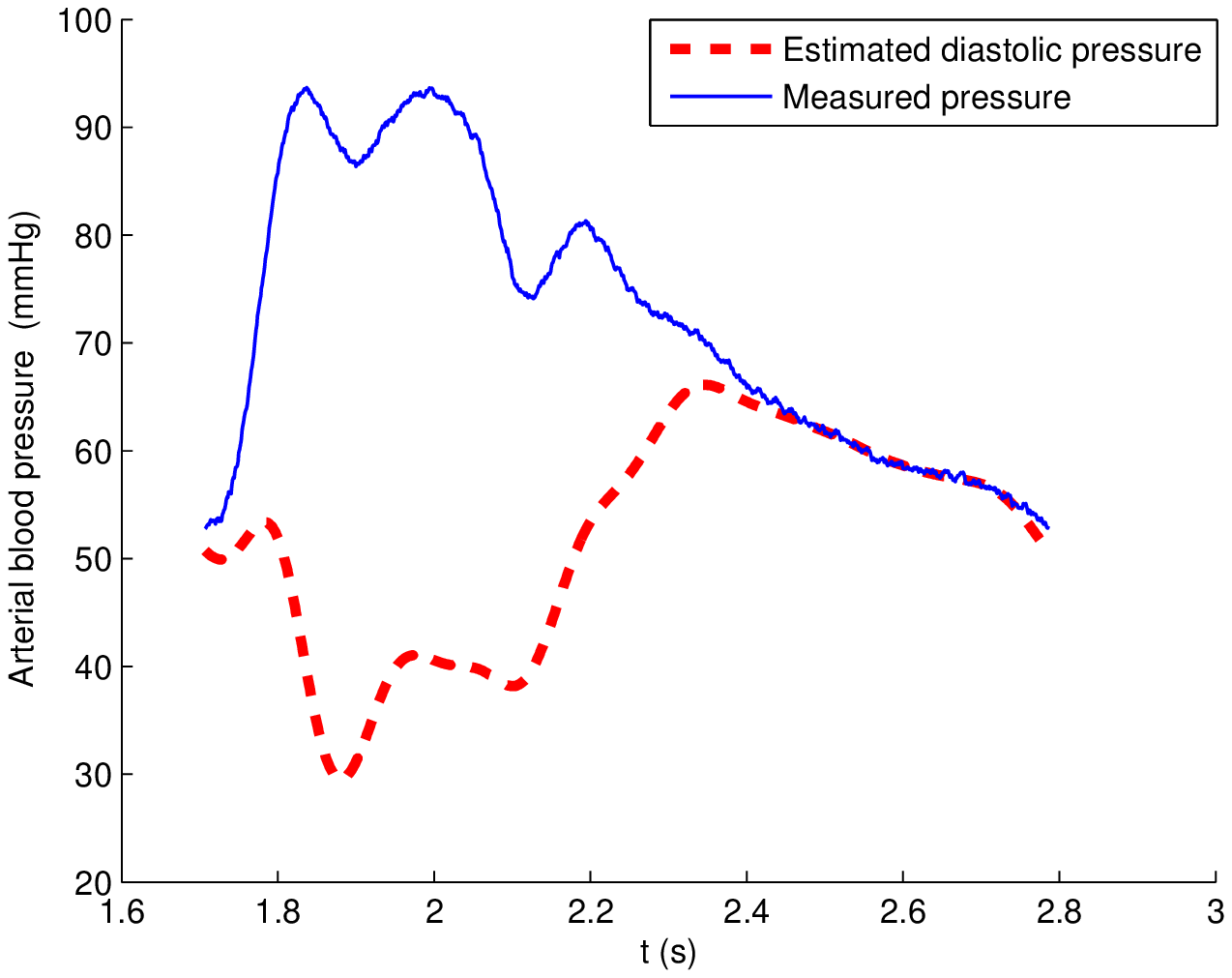,width=6cm}}
 \caption{On the left the estimated systolic pressure. On the right estimated diastolic pressure}\label{soliton_part}
\end{center}\end{figure}

Figure  \ref{soliton_part} represents the  measured pressure and the estimated systolic and diastolic pressures respectively. We notice that $\hat{P}_s$ and
$\hat{P}_d$ are respectively localized during the systole and the diastole.\\

%\begin{center}
%\begin{tabular}{ll}
% \includegraphics[width=6.5cm]{Figures/pression_aorte_systole_1bat} & \includegraphics[width=6.5cm]{Figures/pression_aorte_diastole_1bat}\\
% \hspace{1cm}{Estimated systolic pressure} & \hspace{1cm}{Estimated diastolic pressure}
%\end{tabular}
%\end{center}

%\begin{figure}[htbp]
% \begin{center}
% \includegraphics[width=10cm]{Figures/pression_aorte_systole_1bat}
%  \caption{Estimated systolic pressure}\label{soliton_part}
%\end{center}
%\end{figure}
%
%\begin{figure}[htbp]
%  \begin{center}
%\includegraphics[width=10cm]{Figures/pression_aorte_diastole_1bat}
%  \caption{Estimated diastolic pressure}\label{WK_part}
%\end{center}
%\end{figure}

%We point out that the SCSA introduces some interesting  parameters that give relevant physiological information. These parameters are the negative eigenvalues and the invariants that consist in the momentums of $\kappa_{n\chi}$, $n=1,\cdots,N_\chi$ introduced in proposition \ref{convergence invariants chapitre3}. For example, these new cardiovascular indices allow the discrimination between healthy patients and heart failure subjects \cite{LaMeCoSo:07}. A recent study \cite{LaMePaCoVa:10} shows that the first systolic momentum (associated to the systolic phase) gives information on stroke volume variation.

%---------------------------------------------------------------------------------------------------------------

\section{Discussion}
The spectral analysis of the Schr\"{o}dinger operator introduces two inverse problems:
an inverse  spectral problem and an inverse scattering problem.

On the one hand, the inverse spectral problem aims at reconstructing the potential of
a Schr\"{o}dinger operator with its spectrum (spectral function). It has been extensively studied for instance by Borg,
Gel'Fand, Levitan and Marchenko \cite{GeLe:55} or more recently by Ramm \cite{Ram:98}.  They  considered the half-line case and used two spectra
of the Schr\"{o}dinger operator with two different boundary conditions in order to reconstruct the potential. The inverse spectral problem for a semi-classical Schr\"{o}dinger operator $H_h(y)$ have been recently considered for example by Colin de Verdi\'ere \cite{Colin:08} who proposed to reconstruct the potential locally with a single spectrum or Guillemin and Uribe \cite{GuUr:05} who showed that under some assumptions, the low-lying eigenvalues of the operator determine the Taylor series of the potential at the minimum.

In this work we have studied an inverse spectral approach that is different from classical inverse spectral problems.
Indeed, we used more information to reconstruct the potential by including the eigenfunctions as illustrated by equation (\ref{formule introduction}).

On the other hand, the inverse scattering problem aims at recovering the potential from the scattering data (see appendix A).
Many studies considered this question for instance those of Marchenko \cite{Mar:86} who proved that under some conditions on the scattering data,
a potential in $L_1^1(\mathbb{R})$ can be reconstructed from these scattering data and gave an algorithm for recovering the potential. We can also quote works of Faddeev \cite{Fad:64}, Deift and Trubowitz \cite{DeTr:79} considering potentials in $L_1^1(\mathbb{R})$, Dubrovin, Matveev and Novikov \cite{DuMaNo:76} for periodic potentials.

The convergence of the SCSA when $h\rightarrow 0$ is not easy to study. Using the Deift-Trubowitz formula (\ref{formule deift-trubowitz}) we have tried to consider this problem. However,
  despite interesting results obtained regarding the convergence of some quantities depending only on the continuous spectrum of the Schr\"{o}dinger operator to zero, we did not
  succeed in finding a  result of convergence for $y_h$. A study supported by semi-classical concepts is now under consideration.
   The latter is based on the generalization of the results of G. Karadzhov \cite{Kar:90}.

The SCSA method has given promising results in  the analysis of arterial blood pressure waveforms. More than a good reconstruction of these signals, the SCSA
 introduces interesting  parameters that give relevant physiological information. These parameters are the negative eigenvalues and the invariants which are given by the momentums of $\kappa_{n\chi}$, $n=1,\cdots,N_\chi$ introduced in proposition \ref{convergence invariants chapitre3}. For example, these new cardiovascular indices
 allow the discrimination between healthy patients and heart failure subjects \cite{LaMeCoSo:07}. A recent study \cite{LaMePaCoVa:10} shows that the first systolic momentum (associated to the systolic phase) gives information on stroke volume variation, a physiological parameter of great interest.

We have seen in proposition \ref{approximation exacte} that for a fixed value of $h$ the SCSA coincides with a reflectionless potentials approximation. This point seems to be an interesting avenue of research. Indeed, thanks to the relation between reflectionless potentials and solitons, the approximation by reflectionless potentials could have interesting applications in signal analysis and in particular in data compression. As we said in the previous section, solitons are reflectionless potentials of the Schr\"{o}dinger operator. Gardner et al \cite{GaGrKrMi:74} showed that an $N$-soliton is completely determined by the discrete scattering data and in particular by $2N$ parameters which are the negative
eigenvalues and the normalizing constants. Hence, if $-\chi y$ is an $N_\chi$-soliton, it is given by the following formula:
\begin{equation}\label{yschi1} %{pot3}
y(x)=\frac{2}{\chi}\frac{\partial^2}{\partial
x^2}\ln{(\det{(I+A_\chi)}})(x), \quad x\in \mathbb{R},
\end{equation}
$A_\chi(x)$ is an $N_\chi \! \times \! N_\chi$ matrix of
coefficients
\begin{equation}\label{yschi2}
A_\chi(x)=\left[\frac{c_{m\chi} c_{n\chi}}{\kappa_{m\chi} \!+\!
\kappa_{n\chi}} e^{-(\kappa_{m\chi} \!+\!
\kappa_{n\chi})x}\right]_{n,m},\quad n,m=1,\cdots,N_\chi,
\end{equation}

where $-\kappa_{n\chi}^2$ and $c_{n\chi}$, $n=1,\cdots,N_\chi$  are the negative eigenvalues and the normalizing constants of $H_1(\chi y )$.
Hence, this formula provides a parsimonious representation of a signal.

The convergence of the approximation by solitons (or reflectionless potentials) when
$\chi \rightarrow +\infty$ (equivalently $h\rightarrow 0$) was studied by Lax and Levermore
\cite{LaLe:83} in a different context. Indeed they studied the
small dispersion limit of the KdV equation and approached the
initial condition of the KdV equation by an $N$-soliton that
depends on the small dispersion parameter which is in our case $h$. They showed the results in the mono-well
potential case and affirmed without prove that the result remain
still true for multi-well potentials. However the main limitation of this approach is the difficulty to compute the normalizing constants. This difficulty can be explained by the fact
that these constants are defined at infinity which can not be handled in the numerical implementation. At the best of our knowledge there is no study enabling  the computation
of these parameters apart a recent attempt by Sorine \emph{et al} \cite{SoZhLaCr:08}.

\section{Conclusion}
A new method for signal analysis based on a semi-classical approach  has been proposed in this study: the signal is considered as a potential of a Schr\"{o}dinger operator and then
represented using the discrete spectrum of this operator. Some spectral parameters are then computed leading to a new approach for signal analysis.
This study is a first step in the validation of the SCSA. Indeed, we have assessed here the ability of the SCSA to reconstruct some signals.
We have studied particularly  a challenging application which is the analysis of the arterial blood pressure waveforms. The SCSA introduces a novel approach for arterial blood pressure waveform analysis and enables the estimation of relevant physiological parameters.
A theoretical study is now under consideration regarding the convergence of the SCSA for $h\rightarrow 0$.
The work must be orientated at a second step to the comparison between the performance of the SCSA and other signal analysis methods like Fourier transform or the wavelets and also
to the generalization of the SCSA to other fields.

%-----------------------------------------------------------------------------------------------------------

\section*{Acknowledgments}
The authors thank Doctor Yves Papelier from the Hospital B\'ecl\`ere in Clamart  for providing us arterial blood pressure data.

\appendix

\section{Direct and inverse scattering transforms}
These appendices recall some known concepts on direct and inverse scattering transforms of a one-dimensional Schr\"{o}dinger operator.
For more details, the reader can refer to the large number of references on this subject for instance \cite{AkKl:01}, \cite{CaDe:82}, \cite{DeTr:79}, \cite{EcVa:83}, \cite{Fad:64}.

%----------------------------------------------------------------------------------------------------
We consider here the spectral problem of a Schr\"{o}dinger operator $H_1(-V)$, given by
\begin{equation}\label{schannexe}
-\frac{d^2 \psi }{dx^2}+ V(x,t) \psi = k^2 \psi,\quad k\in
\overline{\mathbb{C}}^+,\quad x\in\mathbb{R},
\end{equation}
where the potential  $V$ such that $V\in \mathcal{B}$.
%satisfies the following hypothesis
%\begin{equation}\label{*annexe}
%V\in L_1^1(\mathbb{R}),\quad \quad \quad  \frac{\partial^m
%}{\partial x^m}V\in L^1(\mathbb{R}), \quad m=1,2.
%\end{equation}
%with \begin{equation} L_1^1(\mathbb{R})=
%\{V | \int_{-\infty}^{+\infty}{|V(x)|(1+|x|) dx}<\infty\}.
%\end{equation}
For simplicity, we will omit the indice $1$ of the spectral parameters in the following.

For $k^2 > 0 $, we introduce the solutions $\psi_{\pm}$ of
equation (\ref{schannexe}) such that
\begin{eqnarray}\label{SPECTREC1_contexte}
    \psi_{-}(k,x)&=&\left\{%
\begin{array}{lc}
  T(k)e^{-ikx} &\hspace{0.5cm} x\rightarrow -\infty, \\
  e^{-ikx} + R_r(k)e^{+ikx} &\hspace{0.5cm} x\rightarrow +\infty, \\
\end{array}%
\right.\\
    \psi_+(k,x)&=&\left\{%
\begin{array}{lc}
  T(k)e^{+ikx} &\hspace{0.5cm} x\rightarrow +\infty, \\
  e^{+ikx} + R_l(k)e^{-ikx} &\hspace{0.5cm} x\rightarrow -\infty, \\
\end{array}%
\right.
\end{eqnarray}
where $T(k)$ is called the transmission coefficient and
$R_{l(r)}(k)$ are the reflection coefficients from the left and
the right respectively. The solution $\psi_-$ for example
describes the scattering phenomenon for a wave  $e^{-ikx}$ of
amplitude $1$, sent from $+\infty$. This wave hit an obstacle
which is the potential so that a part of the wave is transmitted
$T(k) e^{-ikx}$ and the other part is reflected  $R_r(k) e^{+ikx}$. $\psi_+$ describes
the scattering phenomenon for a wave $e^{+ikx}$ sent from  $-\infty$.\\
%\begin{figure}
%\begin{center}
%  \includegraphics[width=8cm]{Figures/ondes}
%  \caption{Phénomène scattering de l'opérateur de Schr\"{o}dinger}\label{ondes2}
%\end{center}
%\end{figure}

For $k^2 <0$, the Schr\"{o}dinger operator spectrum has $N$ negative
eigenvalues denoted  $-\kappa_n^2$, $n=1,\cdots, N$. The associated
$L^2$-normalized eigenfunctions  are such that
\begin{eqnarray}
    \psi_n(x)&=& c_{ln} e^{-\kappa_n x}, \quad \quad \quad \quad \quad x\rightarrow
    +\infty,\\
    \psi_n(x)&=& (-1)^{N-n} c_{rn} e^{+\kappa_n x}, \quad \:\: \: x\rightarrow -\infty,
\end{eqnarray}
$c_{ln}$ and $c_{rn}$ are the normalizing constants
from the left and the right respectively.

The spectral analysis of the Schr\"{o}dinger operator introduces two
transforms:
\begin{itemize}
    \item The direct scattering transform (DST) which consists
    in determining the so called scattering data for a given potential. Let us denote $\mathcal{S}_l(V)$ and
    $\mathcal{S}_r(V)$ the scattering data from the left and the right
    respectively:
\begin{equation}
\mathcal{S}_j(V) : = \{R_{\overline{j}}(k), \; \kappa_n, \;
c_{jn}, \;\;
 n=1,\cdots,N \},\quad \quad j=l,r,
 \end{equation}
where $\overline{j}=r$ if $j=l$ and $\overline{j}=l$ if
$j=r$.

\item The inverse scattering transform (IST) that aims at  reconstructing a potential $V$ using the scattering data.
\end{itemize}

The scattering transforms have been proposed to solve some partial
derivative equations for instance the KdV equation \cite{GaGrKrMi:74}.

\section{Reflectionless potentials}

Deift and Trubowitz \cite{DeTr:79} showed that when the
Schr\"{o}dinger operator potential $V$ satisfies hypothesis
(\ref{hypotheses}) then it can be reconstructed using an explicit
formula given by
\begin{equation}\label{formule deift-trubowitz}
    V(x)=-4\sum_{n=1}^N{\kappa_n\psi_n^2(x)}+\frac{2i}{\pi}\int_{-\infty}^{+\infty}{kR_{r(l)}(k) \: f_{\pm}^2(k,x)
    dk},\quad x\in \mathbb{R}.
\end{equation}
This formula is called the \emph{Deift-Trubowitz trace
formula}. It is given by the sum of two terms: a sum of $\kappa_n
\psi_n^2$ that characterizes the discrete spectrum, and an integral
term that characterizes the contribution of the continuous
spectrum.

There is a special classe of potentials called reflectionless
potentials for which the problem is simplified. A reflectionless
potential is defined by $R_{l(r)}(k) = 0$, $\forall k\in
\mathbb{R}$. According to the Deift-Trubowitz formula, a
reflectionless potential can be written using the discrete spectrum
only,
\begin{equation}\label{expression potentiel sans reflexion}
    V(x,t)= - 4 \sum_{n=1}^{N}{\kappa_n\psi_n^2(x,t)},
\end{equation}
%
%where $-\kappa_n^2$, $n=1,\cdots,N$ are the negative eigenvalues
%of $H_1(-V)$ and $\psi_n$, $n=1,\cdots, N$
%the associated $L^2$-normalized eigenfunctions.

%-----------------------------------------------------------------------------------------------------------
\section{An infinite number of invariants}
%-----------------------------------------------------------------------------------------------------------
There is an infinite number of time invariants for the KdV equation given by the conserved quantities \cite{GaGrKrMi:74}, \cite{GeHo:94}, \cite{MoNoVa:01}. Let us denote these invariants $ I_m(V)$,
$m=0,1,2,\cdots$. They are of the form
\begin{equation}\label{gen1}
I_m(V)=(-1)^{m+1}\frac{2m+1}{2^{2m+2}}\!\!\int_{-\infty}^{+\infty}{P_m(V,\frac{\partial
V}{\partial x},\frac{\partial^2 V}{\partial x^2}, \cdots)dx},
\end{equation}
where $P_m$, $m=0,1,2,\cdots$ are known polynomials in $V$ and its
successive derivatives with respect to $x\in\mathbb{R}$
\cite{CaDe:82}.

A general formula relates  $I_m(V)$ to the scattering data of
$H_1(-V)$ \cite{CaDe:82}, \cite{GeHo:94}, \cite{MoNoVa:01} as follows:
\begin{eqnarray}\label{invgeneral}
I_m(V) = \sum_{n=1}^{N}
    {\kappa_{n}^{2m+1}}+\frac{2m+1}{2\pi}\int_{-\infty}^{+\infty}{{(-k^2)^m \ln{(1-|R_{r(l)}(k)|^2)}dk}},
\end{eqnarray}
$ m=0,1,2,\cdots$. So, for $m=0$, $P_0(V,\cdots)=V$, we get with
(\ref{gen1}) and (\ref{invgeneral}):
\begin{equation}\label{firstinvariant}
 \int_{-\infty}^{+\infty}{V(x) dx}= -4\sum_{n=1}^{N}{\kappa_{n}}
 - \frac{1}{\pi} \int_{-\infty}^{+\infty}{\ln{(1-|R_{r(l)}(k)|^2)}
dk}.
\end{equation}

%\section*{References}

%\bibliography{references}

\begin{thebibliography}{10}

\bibitem{AkKl:01}
{\sc T.~Aktosun and M.~Klaus}, {\em Inverse theory: problem on the line},
  Academic Press, London, 2001, ch.~2.2.4, pp.~770--785.

\bibitem{BlSt:96}
{\sc Ph. Blanchard and J.~Stubbe}, {\em Bound states for {S}chrödinger
  {H}amiltonians: phase space methods and applications}, Rev. Math. Phys., 35
  (1996), pp.~504--547.

\bibitem{CaDe:82}
{\sc F.~Calogero and A.~Degasperis}, {\em Spectral Transform and Solitons},
  North Holland, 1982.

\bibitem{CrSo:07}
{\sc E.~Cr\'epeau and M.~Sorine}, {\em A reduced model of pulsatile flow in an
  arterial compartment}, Chaos Solitons \& Fractals, 34 (2007), pp.~594--605.

\bibitem{Colin:08}
{\sc Y.~Colin de~Verdière}, {\em A semi-classical inverse problem ii:
  Reconstruction of the potential}, Preprint,  (2008).

\bibitem{DeTr:79}
{\sc P.~A. Deift and E.~Trubowitz}, {\em Inverse scattering on the line},
  Communications on Pure and Applied Mathematics, XXXII (1979), pp.~121--251.

\bibitem{DuMaNo:76}
{\sc B.~A Dubrovin, V.~B. Matveev, and S.~P. Novikov}, {\em Nonlinear equations
  of {K}orteweg-de {V}ries type, finite-zone linear operators, and {A}belian
  varieties}, Russian Math. Surveys, 31 (1976), pp.~59--146.

\bibitem{EcVa:83}
{\sc W.~Eckhaus and A.~Vanharten}, {\em The Inverse Scattering Transformation
  and the Theory of Solitons}, North-Holland, 1983.

\bibitem{Fad:64}
{\sc L.~D. Faddeev}, {\em Properties of the {S}-matrix of the one-dimensional
  {S}chrödinger equation}, Trudy Mat. Inst. Steklov, 73 (1964), pp.~314--336.

\bibitem{GaGrKrMi:74}
{\sc C.~S. Gardner, J.~M. Greene, M.~D. Kruskal, and R.~M. Miura}, {\em
  Korteweg-de {V}ries equation and generalizations {VI}. {M}ethods for exact
  solution}, in Communications on Pure and Applied Mathematics, vol.~XXVII,
  J.Wiley \& {S}ons, 1974, pp.~97--133.

\bibitem{GeLe:55}
{\sc I.~M. Gel'fand and B.~M. Levitan}, {\em On the determination of a
  differential equation from its spectral function}, Amer. Math. Soc. Transl.,
  2 (1955), pp.~253--304.

\bibitem{GeHo:94}
{\sc F.~Gesztesy and H.~Holden}, {\em Trace formulas and conservation laws for
  nonlinear evolution equations}, Reviews in Mathematical Physics, 6 (1994),
  pp.~51--95.

\bibitem{GuUr:05}
{\sc V.~Guillemin and A.~Uribe}, {\em Some inverse spectral results for
  semi-classical schr\"{o}dinger operators}, Preprint,  (2005).

\bibitem{HeRo:90a}
{\sc B.~Helffer and D.~Robert}, {\em Riesz means of bound states and
  semiclassical limit connected with a {L}ieb-{T}hirring's conjecture {I}},
  Asymptotic Analysis, 3 (1990), pp.~91--103.

\bibitem{HuGoOr:84}
{\sc M.Y. Hussaini, D.~Gottlieb, and S.~A. Orszag}, {\em Theory and
  applications of spectral methods}, in Spectral Methods for Partial
  Differential Equations, D.~Gottlieb R.~Voigt and M.~Hussaini, eds., 1984,
  pp.~1--54.

\bibitem{Kar:90}
{\sc G. Karadzhov}, {\em Asymptotique semi-classique uniforme de la fonction
spectrale
 d'op\'erateurs de Schr\"odinger.}
  C.R. Acad. Sci. Paris, t. 310, S\'erie I, p.~99-104 (1990).

\bibitem{LaCrPaSo:07}
{\sc T.~M. Laleg, E.~Cr\'epeau, Y.~Papelier, and M.~Sorine}, {\em Arterial
  blood pressure analysis based on scattering transform {I}}, in Proc. {EMBC},
  Sciences and Technologies for Health, Lyon, France, August 2007.

\bibitem{LaCrSo:07J}
{\sc T.~M. Laleg, E.~Cr\'epeau, and M.~Sorine}, {\em Separation of arterial
  pressure into a nonlinear superposition of solitary waves and a windkessel
  flow}, Biomedical Signal Processing and Control Journal, 2 (2007),
  pp.~163--170.

\bibitem{LaCrSo:07}
\leavevmode\vrule height 2pt depth -1.6pt width 23pt, {\em Travelling-wave
  analysis and identification. {A} scattering theory framework}, in Proc.
  European Control Conference {ECC}, Kos, Greece, July 2007.

\bibitem{LaMePaCoVa:10}
{\sc T.~M. Laleg, C.~M\'edigue, Y.~Papelier, F.~Cottin, and A.~Van de~Louw}, {\em
  Validation of a new method for stroke volume variation assessment: a
  comparison with the picco technique}, to appear in Annals of biomedical engineering.

\bibitem{LaMeCoSo:07}
{\sc T.~M. Laleg, C.~M\'edigue, F.~Cottin, and M.~Sorine}, {\em Arterial blood
  pressure analysis based on scattering transform {II}}, in Proc. {EMBC}, Lyon, France, August 2007.

\bibitem{LaLi:58}
{\sc L.~D. Landau and E.~M. Lifshitz}, {\em Quantum Mechanics: Non-Relativistic
  Theory}, vol.~3, Pergamon Press, 1958.

\bibitem{LaWe:00}
{\sc A.~Laptev and T.~Weidl}, {\em Sharp {L}ieb-{T}hirring inequalities in high
  dimensions}, Acta Mathematica, 184 (2000), pp.~87--111.

\bibitem{LaLe:83}
{\sc P.~D. Lax and C.~D. Levermore}, {\em The small dispersion limit of the
  {K}orteweg-de {V}ries equation {I}, {II}, {III}}, Comm. Pure. \& Appl. Math.,
  36 (1983), pp.~253--290, 571--593, 809--828.

\bibitem{Mar:86}
{\sc V.~A. Marchenko}, {\em Sturm-Liouville operators and applications},
  Birkhäuser, Basel, 1986.

\bibitem{MoNoVa:01}
{\sc S.~Molchanov, M.~Novitskii, and B.~Vainberg}, {\em First {K}d{V} integrals
  and absolutely continuous spectrum for 1-d {S}chrödinger operator}, Commun.
  Math. Phys, 216 (2001), pp.~195--213.

\bibitem{Ram:98}
{\sc A.~G. Ramm}, {\em A new approach to the inverse scattering and spectral
  problems for the strum-liouville equation}, Annal. der Physik, 7 (1998),
  pp.~321--338.

\bibitem{ReSi:78}
{\sc M.~Reed and B.~Simon}, {\em Methods of modern mathematical physics, {IV}.
  {A}nalysis of operators theory}, Academic Press, New York, 1978.

\bibitem{SoZhLaCr:08}
{\sc M.~Sorine, Q.~Zhang, T.~M. Laleg, and E.~Crepeau}, {\em Parsimonious
  representation of signals based on scattering transform}, in IFAC'08, July
  2008.

\bibitem{Tre:00}
{\sc L.~N. Trefethen}, {\em Spectral Methods in Matlab}, SIAM, 2000.

\end{thebibliography}

\end{document}